\documentclass[aps,twocolumn,superscriptaddress,preprintnumbers,showpacs,article,prb]{revtex4-1}

\usepackage[latin9]{inputenc}
\usepackage{color}
\usepackage{verbatim}
\usepackage{amsmath}
\usepackage{amssymb}
\usepackage{graphicx}
\usepackage{esint}
\usepackage{esvect}
\usepackage{braket}
\usepackage{amsmath,bm}
\usepackage{graphicx}
\usepackage{subfigure}
\usepackage{url}
\usepackage{lipsum}
\usepackage{lipsum}
\usepackage[dvipsnames]{xcolor}
\usepackage{hyperref}
\usepackage{etoolbox}
\usepackage{lipsum}
\usepackage{dblfloatfix}

\makeatletter
\@ifundefined{textcolor}{}
{%
 \definecolor{BLACK}{gray}{0}
 \definecolor{WHITE}{gray}{1}
 \definecolor{RED}{rgb}{1,0,0}
 \definecolor{GREEN}{rgb}{0,1,0}
 \definecolor{BLUE}{rgb}{0,0,1}
 \definecolor{CYAN}{cmyk}{1,0,0,0}
 \definecolor{MAGENTA}{cmyk}{0,1,0,0}
 \definecolor{YELLOW}{cmyk}{0,0,1,0}
}

\usepackage{mathrsfs}

\draft
\makeatother
\begin{document}

\title{Boundary edge networks induced by bulk topology}
\author{Yan-Qi Wang}
\affiliation{Department of Physics, University of California, Berkeley, Berkeley CA 94720, USA}
\affiliation{Materials Sciences Division, Lawrence Berkeley National Laboratory, Berkeley CA 94720, USA}
\author{Joel E. Moore}
\affiliation{Department of Physics, University of California, Berkeley, Berkeley CA 94720, USA}
\affiliation{Materials Sciences Division, Lawrence Berkeley National Laboratory, Berkeley CA 94720, USA}

\begin{abstract}
We introduce an effective edge network theory to characterize the boundary topology of coupled edge states generated from various types of topological insulators.  Two examples  studied are a two-dimensional second-order topological insulator and three-dimensional topological fullerenes, which involve multi-leg junctions.  As a consequence of bulk-edge correspondence, these edge networks can faithfully predict properties such as the energy and fractional charge related to the bound states (edge solitons) in the aforementioned systems, including several aspects that were previously complicated or obscure.
\end{abstract}

\maketitle
\section{Introduction}
A central feature of topological insulators (TI) is the bulk-edge correspondence: a $d$-dimensional TI with given symmetries has a bulk energy gap but symmetry protected gapless $d-1$ dimensional boundary excitations~\cite{Kane2005,Fu2007Mar,moore&balents-2006,Fu2007July,Qi2011,Hasan2010,Moore2010}. Recent studies on higher-order TIs generalized this bulk-edge correspondence.  An $n$-th order TI has protected gapless modes of co-dimension $n$~\cite{Benalcazar61,Benalcazar2017,Langbehn2017,Ezawa2018,Schindler2018,Song2017,Khalaf2017,Khalaf2018,Fang2017}.  A two-dimensional (2d) second order topological insulator (2d SOTI), for instance, is an insulator with gapped edge but gapless corners~\cite{Benalcazar61,Benalcazar2017,Langbehn2017,Ezawa2018}, i.e., there are localized in-gap states at corners under open boundary conditions. The higher order TIs can be derived from gapping out boundary Hamiltonian~\cite{Langbehn2017, Khalaf2017,Khalaf2018,Fang2017}. More specifically, to obtain a 2d SOTI, one can gap out a single helical edge state~\cite{Benalcazar61,Benalcazar2017,Langbehn2017}, or alternately a pair of coupled counter-propagating helical edge states~\cite{Langbehn2017,Zhu2018,Yan2018}. The point of this paper is to develop an effective theory to describe coupled edge states more generally and their dependence on the topology of the system boundary, which allows a description of the domain-wall states that remain at the intersection of edges for various types of edge junctions.

Meanwhile, one can think of the connected problem of higher order TIs.  If we put an ordinary 2d TI on a closed surface of some 3d manifold, is it possible to have gapped 2d faces and 1d edges, but gapless 0d corner modes? Topological fullerenes~\cite{Ruegg2013pra} are an example of this kind of system. They are polyhedral surfaces wrapped by the Haldane honeycomb lattice model~\cite{Haldane1988}, leaving wedge disclination defects at the vertices~\cite{Ruegg2013prl,Ruegg2013pra}.  While these fullerenes do not currently exist in nature, very recent experiments indicate that twisted bilayer graphene at small twist angle supports a network of domain walls with threefold junctions (``Y-junctions'')~\cite{Huang2018,Wu2018}.  These domain walls~\cite{fengwang} are not strictly topologically protected but conductance is expected to be high at the length scales of this network.  If the planar system has non-vanishing Chern number, these topological fullerenes have gapped bulk and hinge states (here a ``hinge state'' is localized at the intersection of two 2d surfaces), but characteristic corner-localized in-gap states. These corner states can be related to the existence of nontrivial defect states bound to isolated wedge disclinations~\cite{Teo2013,Gopalakrishnan2013,Benalcazar2014}. The connection between the fullerene problem and the 2d SOTI can be viewed as follows: the classification of 2d SOTI is derived from that of TIs in 1d, which is identical to the classification of co-dimension 2 topological defects~\cite{Teo2010,Chiu2016,Langbehn2017}. This implies that the topological fullerenes and certain classes of 2d SOTI should be describable in the same framework.  The emergence of states bound to defects (such as disclination or dislocation) has previously been explained in several cases by edge soliton theory, i.e., the effective theory for a pair of coupled counter-propagating helical edge states~\cite{Lee2007,Qi2008,Ran2009,Ruegg2013prl,Klinovaja2015}.  Although this theory is able to predict the fractional charge bound to the (edge) soliton~\cite{Su1980,Goldstone1981,Jackiw1983} in those examples, one needs to extend the approach in order to incorporate crystalline symmetries in more complicated systems and obtain faithful bound state energies.  (Note that in a system of noninteracting fermions, fractional charge should be thought of as an offset or displacement of the charge density, rather than as a property of elementary excitations.)

In this article, we propose a generic edge network theory to capture the boundary topology of coupled edge states.  As a consequence of the bulk-edge correspondence, the edge states carry the necessary information of their topological insulator parents.  By assigning proper boundary conditions on edge states at their vertices, the edge networks correctly predict the existence of bound states (edge solitons) and other information.  We further considered edge states living on the hinges of varies 3d manifolds, where the edge states are generated from topological insulators attached on corresponding surfaces.  Such edge networks can faithfully predict the energy and fractional charge of bound states located at the vertices, going beyond previous edge soliton theories.  These edge networks are shown here to capture the key properties of topological fullerenes as well as some 2d SOTIs, and it is hoped that they will be useful for other problems as well.

The rest of the paper is organized as follows: In Sec. \ref{Description}, we briefly review basic facts and notation for an edge network made from multiple pairs of coupled helical edge states. In Sec. \ref{minimal}, we discuss the minimal edge network constrained to lie on a closed 1d loop, and show the existence of bound state with fractional charge in the presence of certain symmetries. Based on this we further propose an A\uppercase\expandafter{\romannumeral 1} class 2d SOTI that can be easily realized with atoms in an optical lattice. In Sec. \ref{Multi}, we consider edge networks with a multi-leg vertex. We first derive the bound state energy and charge for a Y-junction via a scattering matrix approach in Sec. \ref{ScatteringMatrix}. Then, in Sec. \ref{Application}, we apply the results in Sec. \ref{ScatteringMatrix} to the tetrahedral topological fullerene as an example. Starting from edge networks, we map the tetrahedral topological fullerene to the 2d SOTI we proposed.  We summarize the main results in Sec. \ref{Conclusion} with an eye toward future developments and applications of this picture.

\section{Description of edge network} \label{Description}
We start with several pairs of coupled helical edge states, e.g., living on the hinges of the 3d manifold shown in Fig.[\ref{EdgeNetWorks}.(a)].  The network is described by the effective Hamiltonian:
\begin{equation}\label{EdgeHam}
	H_{\text{edge}} = \sum_{i}\int dx_{i} \Psi^\dagger (x_{i})(-i v_i\partial_{x_{i}} \sigma_z + {\mathcal M}_i(\theta_i)) \Psi(x_i).
\end{equation}
Here, $i$ labels the hinges, and $x_i$ is the coordinate set along a specific hinge. The two component wave-function $\Psi(x_{i}) = (\psi_\alpha(x_{i}), \psi_\beta(x_{i}))^T$ denotes a pair of coupled counter-propagating helical edge states living on $i$-th hinge, and varies smoothly on the scale of the lattice constant. The magnitude of edge velocity $v$ is set identical for all edge states, and their directions should be compatible with the positive direction of $x_i$. The mass term ${\mathcal M_i(\theta_i)} = m\cos \theta_i \sigma_x + m \sin \theta_i \sigma_y$ describes the coupling on hinge $x_i$, where $\sigma_{x,y,z}$ are Pauli matrixes. Without loss of generality, we assume that $m \geq 0$ and $0\leq \theta_i \leq 2\pi$. If $m =0$, the helical edge states are decoupled and their energy spectrum is gapless.  A non-zero mass term can locally gap out a pair of edge states, which is the situation that we are interested in.

To the Hamiltonian we need to add proper boundary conditions for these edge states at vertices where two or more edges come together. The boundary condition describe the scattering process at the junction. By doing so we can solve Eq.[\ref{EdgeHam}] and predict the existence of localized edge solitons that lie in the (bulk and edge) gaps, as well as their properties.

Before discussing edge network on specific configuration, we point out that the Hamiltonian Eq.[\ref{EdgeHam}] may be generalized into the case of Helical Luttinger liquid ~\cite{Wu2006,Xu2006,Hou2009,Giamarchi2003quantum}:
\begin{equation}\label{HLL}
	\tilde H_{\text{edge}} = \sum_i (H_0^i + H_{\text{int}}^i).
\end{equation}
The noninteracting Hamiltonian $H_0^i$ on each hinge can be divided into two parts: the linearized free Dirac field $H_{0,1}^i$ and their coupling ($H_{0,2}^i$) with two real-valued classical scalar field $\lambda_{1,2}(x_i)$~\cite{Chamon2017topological}:
\begin{equation}
\begin{aligned}
	H_{0,1}^i &=-v \int dx_i {(}\psi^\dagger_{\alpha,i}i\partial_{x_i}\psi_{\alpha,i} - \psi^\dagger_{\beta,i}i\partial_{x_i}\psi_{\beta,i}{)}, \\
	H_{0,2}^i &= \int dx_i(\lambda_{1,i}\psi^\dagger_{\alpha,i} \psi_{\beta,i}+ i\lambda_{2,i}\psi_{\beta,i}^\dagger \psi_{\alpha,i} + \text{H.c.}).
\end{aligned}
\end{equation} 
Here $\psi_{\alpha(\beta),i}$ ($\lambda_{1(2),i}$) is short for $\psi_{\alpha(\beta)}(x_i)$ ($\lambda_{1,2}(x_i)$). Compared with ${\mathcal M}_i(\theta_i)$ in Hamiltonian Eq.[\ref{EdgeHam}], we find that $\lambda_{1,i} = m \cos \theta_i$ and $\lambda_{2,i} = m \sin \theta_i$. For helical Luttinger liquid, we only need to consider the forward scattering $H_{\text{int},2}^i$ and chiral interaction $H_{\text{int},4}^i$~\cite{Wu2006,Xu2006,Hou2009,Giamarchi2003quantum}:
\begin{equation}
	\begin{aligned}
		H_{\text{int,2}}^i&= g_{2,i} \int dx_i (\psi^\dagger_{\alpha,i}\psi_{\alpha,i} \psi^\dagger_{\beta,i} \psi_{\beta,i}),\\
		H_{\text{int,4}}^i&= \frac{g_{4,i}}{2} \int dx_i (\psi^\dagger_{\alpha,i}\psi_{\alpha,i}\psi^\dagger_{\beta,i}\psi_{\beta,i} + \psi^\dagger_{\beta,i}\psi_{\beta,i}\psi^\dagger_{\alpha,i} \psi_{\alpha,i}),
	\end{aligned}
\end{equation}
where $g_{2,i}$ and $g_{4,i}$ are interacting constants. One can conduct the standard bosonization procedure for Hamiltonian Eq.[\ref{HLL}] by defining bosonic field $\partial_{x_i}\tilde \varphi_i = -\pi[\rho_\alpha(x_i)+\rho_\beta(x_i)]$ and $\partial_{x_i}\tilde \theta_i = \pi[\rho_\alpha(x_i)-\rho_\beta(x_i)]$, where $\rho_{\alpha(\beta)}(x_i)$ stands for the density for counter-propagating edge states, i.e. $\rho_{\alpha(\beta)}(x_i)=\psi^\dagger_{\alpha(\beta),i}\psi_{\alpha(\beta),i}$. The $\tilde \theta_i$ here should be distinguished from $\theta_i$ in effective mass ${\mathcal M}$. The Bosonized Hamiltonian for each hinge, $H_B^i= H_{B,0}^i + H_{B,1}^i$ reads:
\begin{equation}
	\begin{aligned}
		H_{B,0}^i &= \frac{1}{2\pi}\int dx_i [uK(\partial_{x_i} \tilde \theta_i)^2 + \frac{u}{K}(\partial_x \tilde \varphi_i)^2],\\
		H_{B,1}^i &= \frac{1}{2\pi a} \int dx_i m \cos (\tilde \varphi_i - \theta_i).
	\end{aligned}
\end{equation}
Here, $u \equiv v\sqrt{(1+g_4/2)^2-(g_2/2)^2}$ is the velocity, $K = \sqrt{(1+g_4/2-g_2/2)/(1+g_4/2+g_2/2)}$ is the Luttinger parameter, and $a$ is the lattice constant whose inverse stands for the momentum cut off of vacuum~\cite{Giamarchi2003quantum,Chamon2017topological,Shankar2017}. The Hamiltonian $H_B^i$ is also interacting, and the interaction $H_{B,1}^i$ can be minimized by set $\tilde \varphi(x_i) = \theta(x_i) + \pi$. Referring to the bonsonized conserved current $j^\mu_i = \epsilon^{\mu\nu}\partial_\nu \tilde \varphi(x_i)/2\pi \approx \epsilon^{\mu\nu}\partial_\nu  \theta(x_i)/2\pi $, for the simplest two terminal junction with two legs $x_{1,2}$ (see in Fig.[\ref{EdgeNetWorks}.(b)]), the topological charge $\hat Q$ is given by~\cite{Chamon2017topological,Frohlich1988}:
\begin{equation}\label{TopoC}
	\hat Q \equiv \int {dx} j_\mu(x) \propto\frac{\epsilon^{01}}{2\pi}[\theta(x_2=+\infty) - \theta(x_1=-\infty)].
\end{equation}
A mass kink of ${\mathcal M}_i(\theta_i)$ implies nonzero topological charge $\hat Q$, see in Fig.[\ref{EdgeNetWorks}.(b)]. This is in accordance with the soliton charge $N_s$ derived from non-interacting Fermionic theory~\cite{Su1980,Goldstone1981,Jackiw1983}, see also Eq.[\ref{1dSol}] in later on Sec. \ref{minimal}. For simplicity, in the rest of our article we will focus on the non-interacting model Eq.[\ref{EdgeHam}]. It is reasonable to believe that the value of soliton charge remains unchanged when turning on interaction because it can be calculated from properties away from the junction. However, the response of bound state energy with respect to external flux may be modified by interaction, and may need a deeper description, e.g., by boundary conformal field theory~\cite{affleckoshikawa,Hou2012}.

\section{Edge states on closed 1d loop} \label{minimal}
\begin{figure}[!t]
\centering 
\includegraphics[width=1\columnwidth]{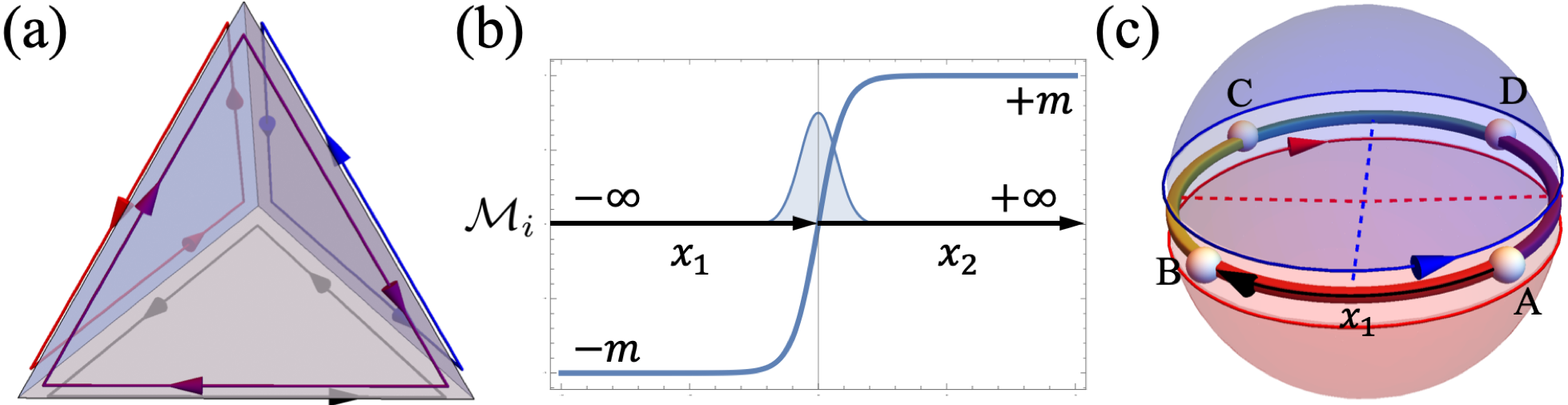}
\caption{\label{EdgeNetWorks} (a) An edge network living on the hinges of a tetrahedron. (b) A mass kink and corresponding soliton in two terminal junction. (c) The minimal edge network. A pair of coupled counter-propagating helical edge states are represented by the blue and red arrows, which can be generated from two Chern insulators with opposite Chern number (see the blue and red hemisphere). The four axes $x_{1,2,3,4}$ are set along the loop in anti clockwise direction, with origins at $A,B,C,D$, respectively.  For simplicity we only plot $x_1$. The red and blue dashed lines stand for two reflection-symmetric axes.}
\end{figure}

We first consider the minimal example of an edge network, a pair of helical edge states living on the boundary of a closed 1d loop, as shown in Fig.[\ref{EdgeNetWorks}.(c)].  The point is to determine how symmetries fix the free coefficients introduced in the previous discussion.  The basis is chosen as $\Psi(x_{i}) = (\psi_\alpha(x_i), \psi_\beta(x_i))^T$, where $\psi_\alpha(x_i)$ ($\psi_\beta(x_i)$) denotes the chiral edge states propagating in the clockwise (anti-clockwise) direction. We set four coordinates $x_{i=1,2,3,4} \geq 0$, and define $x_5=x_1$. The coupling for edge states on each leg is given by an effective mass ${\mathcal M}_i(\theta_i)$, where we have set $v = m =1$ for simplicity. We use a set of trial wave functions $\Phi(x_i)_{o(e)} = \frac{1}{\sqrt{N_{o(e)}}} \exp (-|(x_i-x_i^{o(e)})\sin \varphi| ) \chi(x_i)_{o(e)}$ to look for bound states localized the origin (o) and end (e) of $i$-th edge, with $\chi(x_i)_{o(e)}=(a_i^{o(e)} , b_i^{o(e)} )^T$. Here $a_{o(e)}, b_{o(e)}$, $\varphi$ and normalization constant $1/\sqrt{N_\pm}$ are coefficients to be determined.

Substituting the trial wave function $\Phi{(x_i)}_{o(e)}$ into Eq.[\ref{EdgeHam}] for each individual edge, we find modes localized at two ends of $i$-th edge. For the states at the origin of $i$-th edge, we have the wave function $\chi(x_i)_{o} =e^{i\delta_i^o} (e^{i(\varphi - \theta_i)}, 1 )^T$ with energy $\epsilon_{i}^o = \cos \varphi$. For the states at the end, we have $\chi(x_i)_e =e^{i\delta^e_i} (e^{-i(\varphi + \theta_i)},1 )^T$ with energy $\epsilon_{i}^e = \cos \varphi$. Here, $\delta_{i}^{o,e}$ are overall phase factors. The wave-function we solved previously should satisfy the boundary condition at the corner, i.e., $\Phi({x_{i+1}\rightarrow x_{i+1}^o})_o = \Phi({x_{i}\rightarrow x_i^e})_e$ and $\epsilon_{i+1}^o = \epsilon_{i}^e$. If $\theta_i = \theta_{i+1}$, the only allowed solution is $\varphi = 0$, which means that the localization length $\xi = 1/|\sin \varphi| \rightarrow \infty$ and no bound state exists. If $\theta_i \neq \theta_{i+1}$, we have a mass kink at the intersection of $i$-th and $i+1$-th edge. The solution corresponds to an un-paired edge soliton~\cite{Lee2007} localized at the intersection, with energy and fractional fermion number~\cite{Su1980,Goldstone1981,Jackiw1983} given by:
\begin{equation}\label{1dSol}
	\varphi = |\theta_{i+1}-\theta_i|/2, ~E = \text{sgn}(\theta_{i+1} - \theta_{i}) \cos \varphi, ~N_{s} = -\frac{\varphi}{\pi}.
\end{equation}
Since we measure the charge with respect to the vacuum, there is a minus sign for the soliton charge $N_{s}$. Eq.[\ref{1dSol}] predicts the existence of a domain wall state for any two adjoint edges. More specifically, the edge soliton derived from the aforementioned effective theory can be used to explain fractional charge in varies systems, such as the bound states induced by magnetic domain wall in the quantum spin hall effect~\cite{Qi2008}, or the localized state bound to 2d disclination (dislocation) defect in topological insulators~\cite{Ruegg2013prl,Ran2009}. 

The minimal edge network can explain the corner states in at least some kinds of 2d SOTI. The 2d SOTIs have gapped bulk and edges, but gapless corners. They can be derived from gapping out topological edge states. Heuristically, one potential way to get a 2d SOTI is by stacking 1d TIs, making the 0d boundaries of these 1d TIs form another set of 1d TIs in the perpendicular direction.  This is one way to obtain the quadrupole insulator~\cite{Benalcazar61,Benalcazar2017}. Alternatively, one can couple a pair of (or more) counter-propagating helical edge states living on the boundary of 2d TI and gap them out. Here we will use the latter picture extensively.  Crystalline symmetries with unitary symmetry operator $U$, such as reflection~\cite{Langbehn2017}, inversion\cite{Khalaf2018} and rotation symmetry~\cite{Song2017,Fang2017,Khalaf2017}, can constrain the distribution of effective mass term ${\mathcal M}_i(\theta_i)$ on the boundary. On the edges compatible with crystalline symmetry, $[{\mathcal M}_i(\theta_i), U]=0$. If two adjoint edges are related by crystalline symmetry with operator $U$, then $U^\dagger {\mathcal M}_{i}(\theta_{i}) U ={\mathcal M}_{i+1}(\theta_{i+1})$. If ${\mathcal M}_{i}(\theta_i) \neq {\mathcal M}_{i+1}(\theta_{i+1})$, a domain wall state emerges at the intersection of two adjoining edges, as demonstrated before. 

Distinct from corner-localized zero modes in a 2d second-order topological superconductor~\cite{Langbehn2017,Zhu2018}, we find that, in the absence of particle-hole symmetry and chiral symmetry~\cite{Langbehn2017,Shiozaki2014,Lau2016,Trifunovic2017}, one can have corner modes with non-zero energy. The system we consider has two reflection-symmetric axes, as shown in Fig.[\ref{EdgeNetWorks}.(c)]. The reflection operator for the red axis is $U_b=\sigma_x$, while the reflection operator for the blue axis is $U_r=\sigma_y$. Edge $AB(x_1)$ and $CD(x_3)$ are reflection symmetric edges for $U_b$, thus the only symmetry-allowed mass term is $\pm\sigma_x$. Similarly edge $AD(x_4)$ and $BC(x_2)$ are reflection symmetric edges for $U_r$, thus the only symmetry-allowed mass term is $\pm\sigma_y$. In summary, the effective mass terms on four edges $x_{1,2,3,4}$ are: 
\begin{equation}\label{2dSOTI}
\begin{aligned}
	&{\mathcal M}_1 (0) = +\sigma_x, \quad {\mathcal M}_2 (\frac{\pi}{2}) = +\sigma_y, \\
	&{\mathcal M}_3 (\pi)=-\sigma_x, \quad {\mathcal M}_4 (\frac{3\pi}{2})=-\sigma_y.
\end{aligned}
\end{equation}
Referring to Eq.[\ref{1dSol}], we find $\varphi = {\pi}/{4},~E = \cos \varphi = 1/\sqrt{2}$, and $N_s = -1/4$ for each corner, corresponding to four edge solitons on the loop.

SOTIs have been claimed to be appear in various systems~\cite{Serra-Garcia2018,Peterson2018,Imhof2017}, including bismuth~\cite{Schindler2018Nature}. Based on recent progress of two-dimensional spin-orbit coupling in cold atom system~\cite{Liu2014,Wu2016}, we provide a feasible experimental proposal of 2d SOTI with edge mass distribution as Eq.[\ref{2dSOTI}]. By stacking two Chern insulator layers with opposite Chern numbers (which can be easily realized in experiments by adding a magnetic field with gradient), the 2d tight-binding Hamiltonian for our 2d SOTI model is:
\begin{eqnarray}\label{tbeqn}
		H &=&-\sum_{\langle\bar{i},\vec{j}\rangle_s}t_{\alpha}(\hat c_{\vec{i}\uparrow s}^{\dag}\hat
c_{\vec{j}\uparrow s}-\hat c_{\vec{i}\downarrow s}^{\dag}\hat
c_{\vec{j}\downarrow s})+\sum_{\langle \vec{i} \rangle_s} m_z^s(\hat n_{\vec{i}\uparrow s}-\hat n_{\vec{i}\downarrow s}) \nonumber \\ &&+ \sum_{\langle \vec i \rangle} (2\lambda \hat c^\dagger_{\vec i, \downarrow,+} c_{\vec i, \uparrow,-} - 2\lambda \hat c^\dagger_{\vec i, \uparrow, -} \hat c_{\vec i,\downarrow,+}) \nonumber \\
&&+\bigr[\sum_{\langle j_x \rangle_s}\big{(}it_{\rm so}(\hat c_{j_x\uparrow}^\dag\hat c_{j_x+1\downarrow}-\hat c_{j_x\uparrow}^\dag\hat c_{j_x-1\downarrow})+{\rm H.c.}\big{)}\bigr] \nonumber  \\ &&+\bigr[\sum_{\langle j_y\rangle_s}t_{\rm so}(\hat c_{j_y\uparrow}^\dag\hat c_{j_y+1\downarrow}-\hat c_{j_y\uparrow}^\dag\hat c_{j_y-1\downarrow})+{\rm H.c.}\bigr].
\end{eqnarray}
Here, $s = \pm$ stands for layer index. The positive $t_{\alpha = x,y}$ and $t_{\text{so}}$ denotes, respectively, the inner-layer spin conserved and spin-flip hopping. The $m_z^s$ represents an effective Zeeman term, with $m_z^+ = m_z$ and $m_z^- = - m_z$, which can be realized by a magnetic field with gradient. The spin-flip hopping $t_{\text{so}}$ and $\lambda$ comes from the spin-orbit coupling induced by effective inner-layer and inter-layer Raman coupling, respectively. Transforming $H$ into the momentum space yields $H = \sum_{k,\sigma, \sigma^\prime} \hat c_{\bf{k},\sigma}^\dagger \mathcal{H}_{\sigma,\sigma^\prime}(k) \hat c_{\bf{k},\sigma^\prime}$, with
\begin{equation}\label{Bloch}
\begin{aligned}
	{\mathcal H}(\vec k)= &2 t_{so} \sin{(k_x)}\tau_1  +2t_{so} \sin{(k_y)} \tau_2 \\
	&+(m_z - 2t_x \cos{k_x} - 2t_y\cos{k_y}) \tau_3 \sigma_3 \\ 
	&+  \lambda \tau_1\sigma_1 + \lambda \tau_2\sigma_2,
\end{aligned}
\end{equation}
where $\bm{\tau}$ and $\bm{\sigma}$ are Pauli matrices in spin space and layer space, respectively. If $\lambda =0$, the Hamiltonian Eq.[\ref{Bloch}] has particle-hole symmetry ${\mathcal P} = \tau_1\sigma_3 {\mathcal K}$, time-reversal symmetry ${\mathcal T} = \tau_2\sigma_2{\mathcal K}$, and chiral symmetry ${\mathcal S} = \tau_3\sigma_1$, where ${\mathcal K}$ stands for complex conjugate. With $|m_z| < 2t_x+2t_y$, the system can be viewed as a robust index spin hall effect~\cite{Zhou2008}. It also has two reflection symmetric axes along $x-,y-$ directions, denoted by operator $\hat U_x =\tau_2 \sigma_2 $, $\hat U_y = \tau_1 \sigma_1 $. A small but non-zero $\lambda$ breaks the chiral and particle hole symmetry, and gaps out the helical edge states from the original index spin hall effect. By projecting the low energy Hamiltonian of Eq.[\ref{Bloch}] into the helical edge states derived from $\lambda =0$, one can get the effective edge Hamiltonian identical to Eq.[\ref{2dSOTI}], leading to the similar set of gapped edges and gapless corners.

\begin{figure}[!h]
\centering 
\includegraphics[width=1\columnwidth]{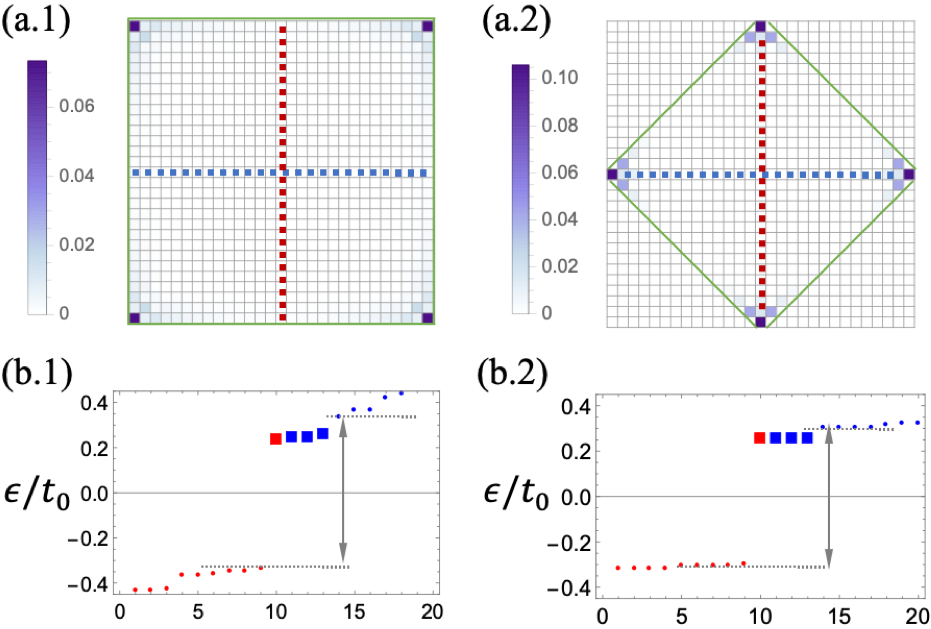}
\caption{\label{SOTI}Numerical results from model Hamiltonian Eq.[\ref{tbeqn}] with two different boundary conditions (marked by green solid lines). (a.1,2) are wave function density for the occupied in gap state, each square stands for one unit cell. The red and blue dashed lines stand for two reflection symmetric axes. (b.1,2) are the energy spectrum close to Fermi surface for the corresponding boundary condition in (a). The squares stand for the corner modes, and the red (blue) stands for the occupied (unoccupied) states at half-filling. The calculations are done with $t_x= t_y = t_0$, $t_{so} = 0.8t_{0}$, $M=0.90t_0$ and $\lambda = 0.3t_0$ for $30 \times 30$ lattice.}
\end{figure}

We further confirm the analytic results by numerically diagonalizing the Hamiltonian Eq.[\ref{tbeqn}] for two different boundary conditions, as shown in Fig.[\ref{SOTI}.(a)]. We find four corner modes with non-zero energy for both patterns. Fig.[\ref{SOTI}.(b)] shows the energy spectrum close to the Fermi surface. The inter-layer coupling $\lambda$ opens a gap $E_{\text{gap}}\approx 2\lambda$ at the boundary, and we can see clearly four corner-localized in gap states. At half-filling, one out of four in-gap states is filled, which compensates the $-1/4$ defect charge at each corner. In realistic cold atom experiments, the detection of the fractional charge at the corner can be conducted by conventional single site resolution. By turning on an $s$-wave onsite interaction for atoms\cite{Liu2014}, this model becomes a 2d second order topological superfluid.

\section{Edge states in multi-leg junction}\label{Multi} 
We now turn to study edge networks with multiple pairs of edge states coming together at a vertex (or equivalently a junction). Fig.[\ref{Tetrahedron}.(a)] shows a Y-junction with six edge states living on three legs. For each semi-infinite axis $x_i(i=1,2,3)$, we use the $\psi_\alpha (x_i)$ and $\psi_\beta (x_i)$ to denote the outgoing and incoming chiral edge states for the $i$-th leg, respectively. Instead of matching wave functions by hand as in the previous minimal 1D edge network, here, we introduce a more generic scattering matrix approach: injecting a mode along a specified leg will lead to reflection and transmission after scattering at the junction, and the poles of scattering matrix implies the existence of bound states. We make the following assumptions to capture the scattering process: (1) Away from the junction, each chiral edge state should be identical to that of an isolated Chern insulator layer, at most up to a global phase factor; (2) During the scattering process, the edge states from the same Chern insulator layer should maintain their amplitude, but could capture a phase shift. The value of the phase shift depends on the details of scattering, but will be constrained by symmetries in specific examples.

\subsection{Scattering theory for Y-junction}\label{ScatteringMatrix}
For an isolated junction, the incoming and outgoing scattering modes can be described by combining incoming and outgoing chiral edge states. Different from localized state, for scattering state with momentum $k$, we denote $\eta = v/m$ and set $k\eta = \sinh \varphi >0$. Then for $x_i > 0$, under the basis $\big{(}\psi_\alpha(x_i),\psi_\beta(x_i)\big{)}^T$, for each individual leg, from Eq.[\ref{EdgeHam}] we can derive normalized wave function of incoming and outgoing modes as: $\psi_{\text{in}}^T(x_i) = (e^{-\varphi -i\theta_i}, 1)^T/\sqrt{1+e^{-2\varphi}},\quad \psi_{\text{out}}^T(x_i) = (e^{\varphi - i\theta_i}, 1)^T/\sqrt{1+e^{+2\varphi}}$, with corresponding energy $E/m = +\cosh \varphi$. If we inject a mode along negative $x_1$ direction, the wave function on leg $x_1$ is given by $\Psi_1(x_1) = e^{-ikx_1}\psi_{\text{in}}(x_1) + r_1 e^{ikx_1} \psi_{\text{out}}(x_1)$. Meanwhile, the wave function on leg $x_{2}$ is given by $\Psi_{2}(x_2) = t_{12} e^{ikx_2} \psi_{\text{out}}(x_2)$, and $\Psi_3(x_3) = t_{13} e^{ikx_3} \psi_{\text{out}}(x_3)$ for wave function on leg $x_3$. We have used $r_1$ for reflection coefficient on leg $x_1$, and $t_{12}~(t_{13})$ for transmission coefficient for the scattering from $x_1$ to $x_2~(x_3)$. With this we can expand the wave function around the intersection as: 
\begin{widetext}
\begin{equation}
\left\{
             \begin{aligned}
             &\Psi_1(0) = \big{(}\frac{e^{-\varphi - i\theta_1}}{\sqrt{1 + e^{-2\varphi}}} + r_1 \frac{e^{\varphi - i\theta_1}}{\sqrt{1+e^{2\varphi}}} \big{)}\psi_{\alpha}({x_1 = 0}) + \big{(} \frac{1}{\sqrt{1 + e^{-2\varphi}}} + \frac{r_1}{\sqrt{1+e^{2\varphi}}} \big{)} \psi_{\beta}({x_1=0}),   \\
             &\Psi_2(0) = t_{12}  \frac{e^{\varphi - i\theta_2}}{\sqrt{1+e^{2\varphi}}} \psi_\alpha({x_2 = 0}) + t_{12}\frac{1}{\sqrt{1+e^{2\varphi}}} \psi_\beta({x_2= 0}), \\
            &\Psi_3(0) = t_{13}  \frac{e^{\varphi - i\theta_3}}{\sqrt{1+e^{2\varphi}}} \psi_\alpha({x_3 = 0}) + t_{13}\frac{1}{\sqrt{1+e^{2\varphi}}} \psi_\beta({x_3= 0})   . 
             \end{aligned}
\right.
\end{equation}
\end{widetext}

\begin{figure}[!h]
\centering 
\includegraphics[width=1\columnwidth]{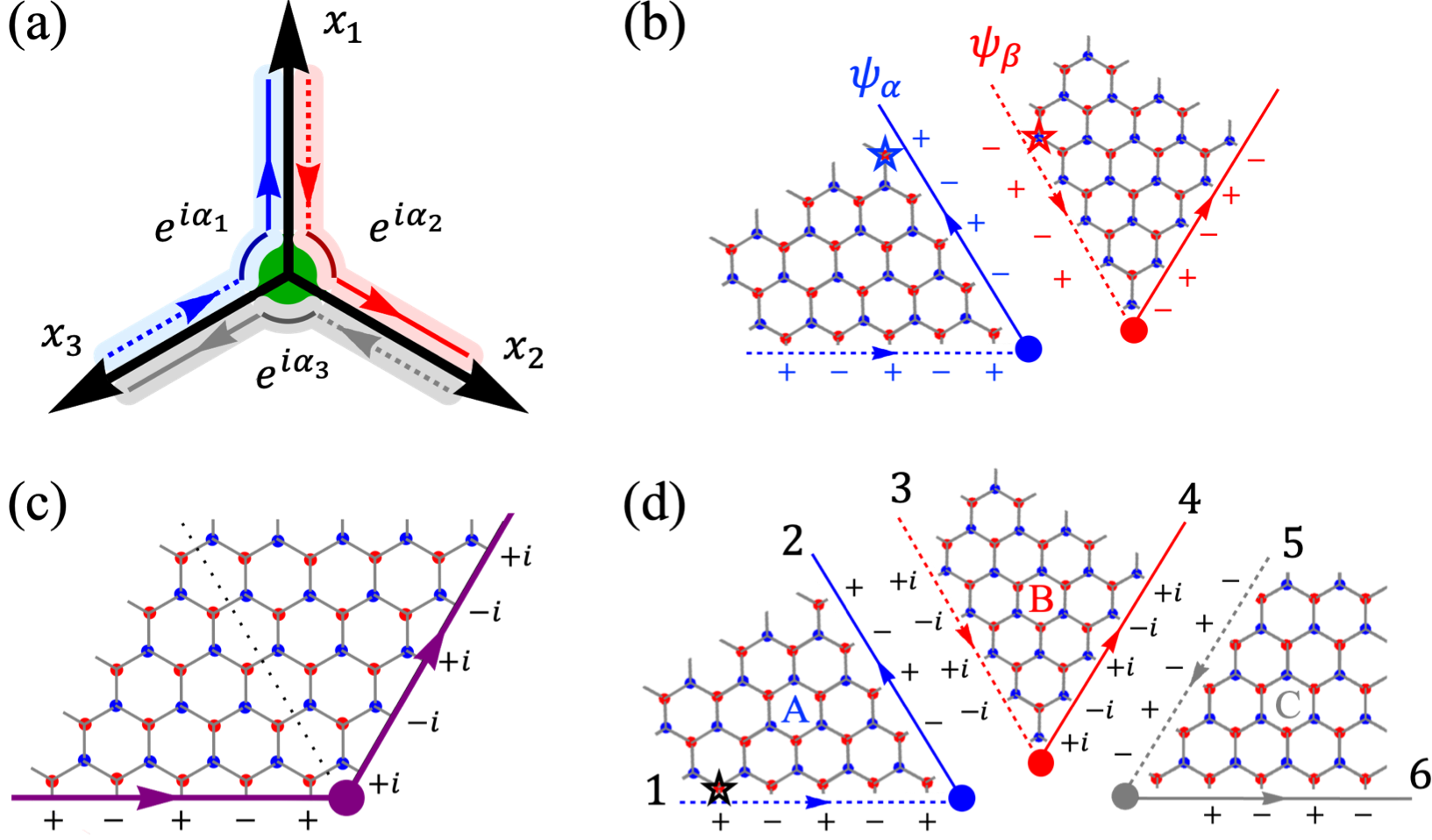}
\caption{\label{Tetrahedron}(a) Edge network for a vertex with three legs (Y-junction). The center of the junction is marked by the green disk. Three coordinates, $x_{1,2,3}$ start from the center and point outward. The solid and dashed arrows in blue, red and gray stands for three pairs of coupled helical edge states. Edge states in the same color are from the same Chern insulator layer. (b) Edge states for two isolated $60^\circ$ Chern insulator slices. The blue (red) $\pm$ stands for the relevant phase factor of edge states measured from $\psi_{\alpha(\beta)}$, with two individual reference points (marked by stars). (c) Edge states for an individual $120^\circ$ Chern insulator slice. (d) A vertex of the type appearing in tetrahedron topological fullerenes and relevant edge states. The $\pm$ stands for the relevant phase factor of edge states measured from edge states $\psi_1$ whose reference point is marked by black star.}
\end{figure}

In Fig.[\ref{Tetrahedron}.(a)], the edge states in same color are from the same Chern insulator layer. Due to the continuity of edge state wave function for each individual layer, we have $\psi_\alpha(x_1 \rightarrow 0^+) = \psi_\beta(x_3 \rightarrow 0^+)$, $\psi_\beta(x_1 \rightarrow 0^+) = \psi_\alpha (x_2 \rightarrow 0^+)$, and $\psi_\beta(x_2 \rightarrow 0^+) = \psi_\alpha (x_3 \rightarrow 0^+)$. During the scattering process they can capture an additional phase factor $e^{i\alpha_i}$, which depends on the details of the scattering process. This leads to:
\begin{equation}
\left\{	
\begin{aligned}
		&\frac{e^{-\varphi - i\theta_1}}{\sqrt{1 + e^{-2\varphi}}} + r_1 \frac{e^{\varphi - i\theta_1}}{\sqrt{1+e^{2\varphi}}} = t_{13}\frac{1}{\sqrt{1+e^{2\varphi}}}e^{i\alpha_1}, \\
		&t_{12} \frac{e^{\varphi - i\theta_2}}{\sqrt{1+e^{2\varphi}}} =\big{(}\frac{1}{\sqrt{1 + e^{-2\varphi}}} + r_1\frac{1}{\sqrt{1+e^{2\varphi}}}\big{)} e^{i\alpha_2}, \\
		&t_{13} \frac{e^{\varphi-i\theta_3}}{\sqrt{1+ e^{2\varphi}}} = t_{12} \frac{1}{\sqrt{1+e^{2\varphi}}}e^{i\alpha_3}.
	\end{aligned}
\right.
\end{equation} 
With this we can solve $r_1$, $t_{12}$ and $t_{13}$ in the term of $\varphi,\alpha_i$, and $\theta_i$. By injecting modes along the negative directions of rest two legs (see in Appendix), we can derive whole coefficients for the scattering matrix $S$:
\begin{equation}\label{Matrix}
	S= \frac{1}{e^{3\varphi} - e^{i\Lambda}} \begin{pmatrix}
		\tilde r_1 & \tilde t_{12} & \tilde t_{13} \\
		\tilde t_{21} & \tilde r_2 & \tilde t_{23} \\
		\tilde t_{31} & \tilde t_{32} & \tilde r_3
	\end{pmatrix},
	~~ \Lambda = \sum_i (\theta_i + \alpha_i).
\end{equation}
For arbitrary scattering process, $\Psi_{\text{out}} = S \Psi_{\text{in}}$, where $\Psi_{\text{in(out)}}^T=\big{(}\psi(x_1),~\psi(x_2),~\psi(x_3)\big{)}_{\text{out(in)}}$. The pole of the scattering matrix, $e^{3\varphi} - e^{i\Lambda} =0$, implies the existence of bound state localized at the junction. Note that, in the presence of edge soliton, each of these semi-infinite legs contributes a fractional charge $-\theta_i/2\pi$. With these we find:
\begin{equation}\label{ScatteringPole}
	\frac{E}{m} = \cosh \varphi = \cos \bigg{(}\frac{\Lambda + 2n\pi}{3}\bigg{)}, ~n \in \mathbb Z,~ N_s = - \frac{\sum_i \theta_i}{2\pi}.
\end{equation}
As we mentioned before, $\Lambda = \sum_i (\theta_i + \alpha_i)$, which depends on the details of scattering.  The energy-phase relation Eq.[\ref{ScatteringPole}] for 3-leg Y-junction can be easily generalized to $l$-leg junction: 
\begin{equation}\label{MultiLeg}
	\frac{E}{m} = \cosh \varphi = \cos \bigg{(}\frac{\Lambda }{l}\bigg{)},~ N_s = - \frac{\sum_i \theta_i}{2\pi},
\end{equation}
where we have let $2n\pi$ be absorbed into $\Lambda$ for latter convenience.

\subsection{Application to topological fullerenes}\label{Application}
The multi-leg edge junction can be used to describe the bound state in an isolated wedge disclination~\cite{Ruegg2013prl,Ruegg2013pra,Teo2013,Gopalakrishnan2013,Benalcazar2014}, which is the building block of topological fullerenes~\cite{Ruegg2013prl,Ruegg2013pra}. More specifically, the Y-junction edge network mentioned above can be used to analyze one vertex of tetrahedral topological fullerenes (as shown in Fig.[\ref{EdgeNetWorks}.(a)]), which is a wedge disclination defect with Frank index $f=3$ (or $180^\circ$ Frank angle). The Frank index $f$ here stands for the number of $60^\circ$ Chern insulator layers taken away from the complete Haldane lattice. In order to build the edge network for such a disclination, let us first consider three $60^\circ$ semi-infinite triangular layers (A,B,C) of Haldane honeycomb lattice coming together, as shown in Fig.[\ref{Tetrahedron}.(d)]. Each layer is coupled with its two neighbors across the seams. The tight-binding Hamiltonian for such a disclination is given by\cite{Haldane1988,Ruegg2013prl,Ruegg2013pra}:
\begin{equation}\label{Haldane}
	{\mathcal H} = -t_0\sum_{\langle i,j \rangle} (c^\dagger_i c_j + \text{H.c.}) - t_1 \sum_{\langle\langle i,j\rangle \rangle} (e^{-i\phi_{ij}}c^\dagger_i c_j +\text{H.c.}).
\end{equation}
Here, $c^\dagger_i$ ($c_i$) is creation (annihilation) operator for spinless fermion on $i$-th site. The $t_0$ and $t_1$ denotes, respectively, the nearest-neighbor hopping and next-nearest-neighbor hopping amplitudes. The $e^{i\phi_{ij}}$ provides an additional phase factor for next-nearest-neighbor hopping. Within the topological region, each individual layer can provide chiral edge states surrounding the bulk. The local Chern vector\cite{Bianco2011} for each layer points outside the plane of the paper, which ensures six edge states propagating according to the pattern in the figure. 

These six edge states are not independent. The blue (red, gray) edge states $1,2$ ($3,4$; $5,6$) come from the same triangular layer, and they are connected by $\psi_{2j}(x_j \rightarrow 0^+) = \psi_{2j-1}(x_{j+2}\rightarrow 0^+)$.  If an edge state is coupled with its time-reversal counterpart across the seam, we say this seam does not have phase mismatch. The total wave function on a lattice site across the seam is given by $\varphi_{\text{edge},\alpha}(x_i) = e^{ik_E a}\psi_\alpha(x_i)$ and $\varphi_{\text{edge},\beta}(x_i) = e^{-ik_Ea} \psi_\beta(x_i)$, respectively, where $k_E$ denotes the edge momentum and $a$ stands for the lattice constant. The $\psi_{\alpha,\beta}(x_i)$ here should be understood as the edge states on corresponding sub lattice. The effective coupling between two states is $\int d\tau \lambda \varphi^*_{\text{edge},\alpha}(x_i)\varphi_{\text{edge},\beta}(x_i)$, with $\lambda$ stands for the bond across the seam. The integral is done within a unit cell. For an isolated disclination, the total phase mismatch $\sum_i \theta_i$ for all legs (seams) is fixed in the absence of external flux. Due to the quantization of charge pumping, the function $\sum_i\alpha_i $ should be linear to $\sum_i \theta_i$, i.e. $\sum_i\alpha_i = A \sum_i \theta_i + B$. The coefficients $A,B$ are related to the parameters from the tight-binding model, such as the effective radius $\rho$ and the Haldane gap $m=3\sqrt{3}t_1$\cite{Ruegg2013prl}. By comparing with the results from exact diagonalizing the tight-binding Hamiltonian Eq.[\ref{Haldane}] (see in Appendix), we find that, for Haldane gap $m \approx t_0=1$,
\begin{equation}\label{Dis180}
	\varphi = \frac{2{\sum_{i}} \theta_i - \pi/2}{3},  ~E = \cos \varphi, ~N_s = -\frac{\sum_i \theta_i}{2\pi}.
\end{equation}
Similarly, for the vertex of an octahedral topological fullerene, the number of legs is $l = 6-2 =4$, and we further have $\varphi = 2\sum_i\theta_i/4$, $E = \cos \varphi$, and $N_s = -\sum_i \theta_i/2\pi$ (with $1 \leq i \leq 4$). For the vertex of an icosahedral topological fullerene, the number of legs is $l=6-1 = 5$, and we further have $\varphi = (2\sum_i \theta_i + \pi/2)/5$, $E = \cos \varphi$, $N_s = -\sum_i \theta_i/2\pi$ (with $1\leq i \leq 5$). 

We now turn to determine the value of $\theta_i$ for each leg\cite{Ruegg2013prl}, especially for the cases with external flux. Let us first consider the process of combining two smaller $60^\circ$-layers in Fig.[\ref{Tetrahedron}.(b)] to a larger $120^\circ$-layer in Fig.[\ref{Tetrahedron}.(c)]. The two smaller layers are cut from the same Haldane honeycomb lattice model, and they are next to each other in the original lattice. With the open boundary condition, both of them can hold chiral edge states, which are denoted by red and blue arrows in Fig.[\ref{Tetrahedron}.(b)]. We can set a simultaneous coordinate for both layers across the seam, thus the total wave function on a lattice site on the blue (red) edge is $\varphi_{\text{edge},\alpha}(\xi) = e^{-ik_E \xi} \psi_{\alpha}(\xi)$ ($\varphi_{\text{edge},\beta} = e^{ik_E\xi} \psi_\beta(\xi)$). In the presence of inversion symmetry, $k_E a =\pi$ for Haldane honeycomb lattice model~\cite{Ruegg2013prl}. Thus the base functions $e^{ik_E\xi}$ oscillates with a period of two sites. In order to glue two layers back to a larger layer without phase mismatch across the seam, the amplitudes should be in the pattern in Fig.[\ref{Tetrahedron}.(b)]. The edge states on the decoupled two branches can be written separately as $\varphi_{\text{edge},\alpha} = e^{-ik_E \xi} \psi_\alpha$ for the lower branch of blue edge, and as $\varphi_{\text{edge},\beta} = e^{ik_E \xi} \psi_\beta$ for the right branch of the red edge. However, as shown in Fig.[\ref{Tetrahedron}.(c)], the edge states has an additional phase shift when bypassing the corner. This leads to $i\psi_\alpha = -\psi_\beta$ or $\psi_\alpha = i\psi_\beta$. Thus we have that the proper phase difference across the seam should be $\pm i$. To avoid any ambiguity induced by the gauge chosen for wave functions, we define the effective mass term ${\mathcal M}_i(\theta_i)$ on each leg with respect to the scenario without phase mismatch. Thus if there is no phase mismatch on a certain leg, then ${\mathcal M}_i(\theta_i=0) = m\sigma_x$.

Note that, for a wedge disclination with Frank index $f=3$ in Fig.[\ref{Tetrahedron}.(d)], if we glue AB and BC across the seam as shown in Fig.[\ref{Tetrahedron}.(b)], the system can be viewed as a Haldane honeycomb lattice on the half plane. The gluing process means that we have chosen to measure the relevant phase factor of edge states on all layers from $\psi_1$ with a fixed reference point. Thus the coupling across the seams $AB$ and $BC$ should not have a phase mismatch, thus ${\mathcal M}_1(0) = {\mathcal M}_2(0) = m\sigma_x$. However, the lower boundaries of A and C has phase mismatch and ${\mathcal M}_{3}(\pi/2) = m \sigma_y$~\cite{Ruegg2013prl}. Finally, referring to Eq.[\ref{Dis180}], we have $\varphi = \pi/6$, $E = \cos \pi/6$ and $N_s = -1/4$ for the vertex of Tetrahedral topological fullerene. Eq.[\ref{Dis180}] also stands in the presence of external flux. Adding an external flux $\Phi$ opposite to local Chern vector at the center of junction is equivalent to change the coupling pattern with additional phase factor $e^{i\Phi}$ for the bond across the Dirac string~\cite{Ruegg2013prl,Ruegg2013pra}. For simplicity we can put the Dirac string along $x_3$, thus $\theta_3 = \phi + \pi/2$ and Eq.[\ref{Dis180}] can be written as $\varphi = 2\Phi/3 + \pi/6$. More specifically, if $\Phi = \pi/2$, we have $\sum_i \theta_i = \pi$ and $\varphi =  \pi/2$. Thus the external flux $\Phi = \pi/2$ moves the bound state energy to $E = \cos \varphi  = 0$, as well as the fractional charge to $N_s = -\sum_i\theta_i/2 = -1/2$. This is consisted with the analysis from symmetry: an external flux with $\Phi = \pi/2$ can restore the particle hole symmetry of the system~\cite{Ruegg2013pra}. Thus the bound state energy should be $0$ and the fractional charge should be $-1/2$. Similar results apply for vertices of octahedral and icosahedral topological fullerenes (see in Appendix), and are in accordance with numerical results~\cite{Ruegg2013prl,Ruegg2013pra}.
\begin{figure}[!t]
\centering 
\includegraphics[width=1\columnwidth]{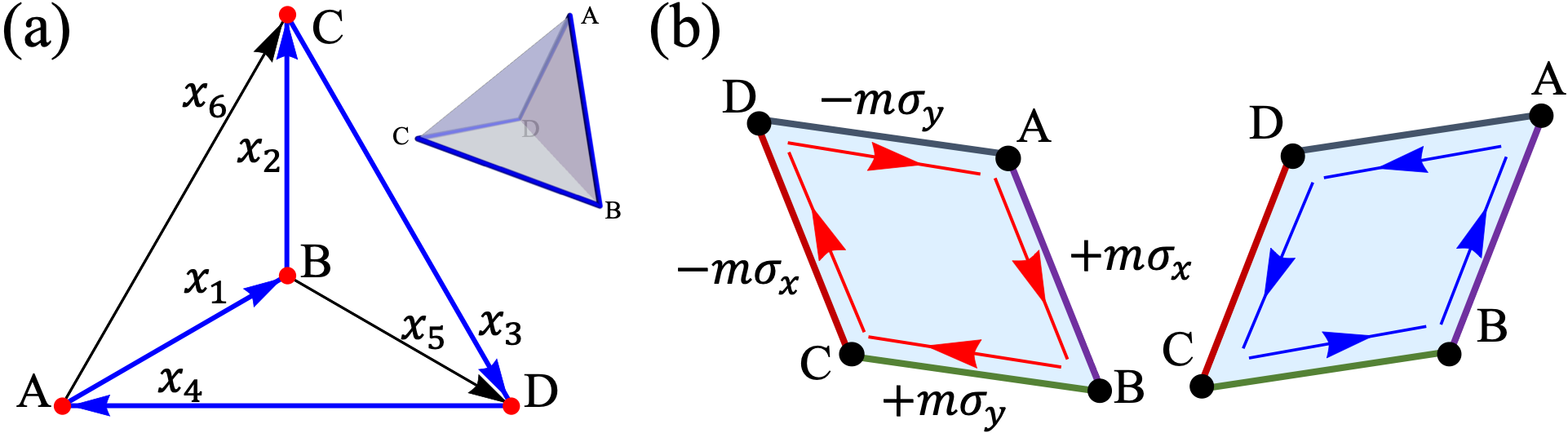}
\caption{\label{Mapping}(a) Edge network and relevant coordinates for tetrahedral topological fullerene. The blue line shows the traversal along the hinges. (b) Mass distribution of edge network for tetrahedral topological fullerene. Cutting the Tetrahedron along the blue line in (a) leads to two parallelograms in (b), which helps to map the tetrahedral topological fullerenes to a 2d SOTI (Eq.[\ref{Bloch}]).}
\end{figure}

The corner states in topological fullerenes can be further explained by the edge networks with a group of multi-leg junctions. In Fig.[\ref{Mapping}.(a)] we plot the edge network for the tetrahedral topological fullerene, with ${\mathcal M}_1(0) = m\sigma_x$, ${\mathcal M}_2(\pi/2) = m\sigma_y$, ${\mathcal M}_3(\pi) = -m\sigma_y$, ${\mathcal M}_4(3\pi/2) = -m\sigma_x$, and ${\mathcal M}_5(0) = {\mathcal M}_6(0) = m\sigma_x$. However, Eq.[\ref{Dis180}] is derived for an isolated vertex with all coordinates point outward, which is slightly different from the settings in Fig.[\ref{Mapping}.(a)]. Note that, for a pair of helical edge states living on a $i$-th hinge with effective mass ${\mathcal M}_i(\theta_i)$, changing the direction of coordinates is equivalent to changing the mass term to $\tilde {\mathcal M}_i\big{(} (-\theta_i)~\text{mod}~2\pi\big{)}$. Thus for each individual vertex, we can first flip the coordinates to the pattern in Fig.[\ref{Tetrahedron}.(a)], by then using Eq.[\ref{Dis180}] we find four corner localized states with $E = \cos \pi/6$ and $N_s = -1/4$.

\section{Conclusion}\label{Conclusion}
We have constructed a generic edge network theory and shown its ability to capture the boundary topology of coupled edge states with different geometric constraints.  We first discussed the minimal edge network on a closed 1d loop, and demonstrated that crystalline symmetry can produce spatial-dependent mass term, leading to the domain wall states at the intersection of adjoint edges.  After discussing a model 2d second-order TI, we constructed edge networks for multi-leg junctions, which can faithfully reflect the properties of bound states  in disclination defects. The edge network can include polyhedral hinges, which allows determination of the corner states in topological fullerenes. These results can help to understand the origin of topologically generated localized states in a variety of situations.

We can view the similarities between the 2D second-order TI and the 3D topological fullerine as reflecting the fact that the classification of 2d SOTI is derived from that of TIs in 1d, which is the same as classification of co-dimension 2 topological defects~\cite{Teo2010,Chiu2016,Langbehn2017}, including point defects in surfaces.  In this sense, the 2d SOTI we proposed is in the same topological class as a corresponding system with wedge disclination defects.  Based on effective edge theory, we can map the topological fullerenes to the 2d SOTI Eq.[\ref{Bloch}] derived from gapping out helical edge states in Sec. \ref{minimal}. For any polyhedron, one can traverse all the vertices along hinges without repeats. The traversal forms a closed 1d loop (see the blue thick arrows in Fig.[\ref{Mapping}.(a)]). We can cut the polyhedron into two congruent Chern insulator layers along the traversal, as shown in Fig.[\ref{Mapping}.(b)]. The two Chern insulator layers can be viewed as a ``twisted'' index spin hall effect. The edges on the closed 1d loop are gapped out by the gluing process, and the effective mass changes after bypassing each corner due to crystalline symmetries, leading to an edge soliton with fractional charge located at the corner. This is identical to the generation of fractional charge in our 2d SOTI model.  Similarly, we can also map the octahedral and icosahedral topological fullerenes to (less natural) 2d SOTIs.

More generally, the networks of edges discussed here could be generalized to incorporate proximity-induced superconductivity or Luttinger liquid corrections, or conceivably to include additional localized degrees of freedom such as boundary Majorana states or spins as in previous studies of the Kondo effect in Y-junctions~\cite{affleckoshikawa}.  In the cases discussed here, there are enough symmetries or other physical constraints to determine the key properties of the localized states in an edge network quite directly, while in other situations the properties such as fractional offset charges might be actively tuned by symmetry-breaking perturbations.  Planar networks of helical edges and three-leg junctions have recently been discovered in bilayer graphene at small twist angles, which suggests that the study of edge networks is likely to become increasingly relevant to experiment.

\section*{\uppercase{Acknowledgements}}
We thank Daniel E. Parker and Takahiro Morimoto for useful conversations. This work was supported as part of the Center for Novel Pathways to Quantum Coherence in Materials, an Energy Frontier Research Center funded by the U.S. Department of Energy, Office of Science, Basic Energy Sciences.  J.E.M. acknowledges additional support from a Simons Investigatorship.

\appendix
\setcounter{figure}{0}
\setcounter{equation}{0}

\renewcommand\theequation{A\arabic{equation}}
\renewcommand\thefigure{A\arabic{figure}}
\section*{\uppercase{Appendix}}
\renewcommand\thesection{\arabic{subsection}}
\renewcommand\thesection{\arabic{subsubsection}}
\subsection{Edge network for two-terminal junction}
In this section, we look into the scattering theory of the simplest two-terminal junction. In order to keep in accordance with the scattering theory in Sec. \ref{ScatteringMatrix}, we set the positive direction of the two legs being opposite to each other and pointing outside the junction. This switches the $\theta_1$ to $-\theta_1$ compared with the notation in Sec. \ref{minimal}.
\begin{figure}[!h]
\centering 
\includegraphics[width=1\columnwidth]{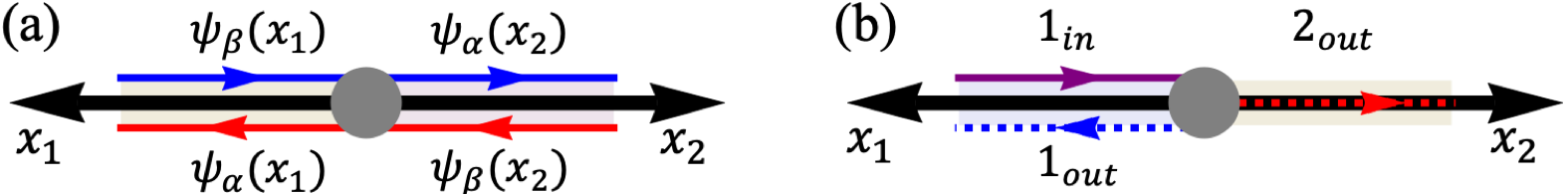}
\caption{\label{1DScattering} 1D Scattering process. (a) Edge network for two terminal junction. (b) The scattering process for two terminal junction with a wave inject along negative $x_1$ direction.}
\end{figure}

We construct the conventional scattering theory as following: suppose we have a wave injected along the negative $x_1$ direction. The wave function on leg $x_1$ is given by $\Psi_1 = e^{-ikx_1}\psi_{1,\text{in}} + r e^{ikx_1} \psi_{1,\text{out}}$. Meanwhile, the wave function on leg $x_2$ is given by $\Psi_2 = t e^{ikx_2} \psi_{2,\text{out}}$. Here $r$ and $t$ stand for the reflection and transmission coefficients, respectively. Different from a localized state, for a scattering state with momentum $k$, we denote $\eta = v/m$ and set $k\eta = \sinh \varphi >0$. Then for $x_i > 0$, using the basis $\big{(}\psi_\alpha(x_i),\psi_\beta(x_i)\big{)}^T$, for each individual leg, from Eq.[\ref{EdgeHam}] we can derive a normalized wave function of incoming and outgoing modes as: $\psi_{\text{in}}^T(x_i) = (e^{-\varphi -i\theta_i}, 1)^T/\sqrt{1+e^{-2\varphi}},\quad \psi_{\text{out}}^T(x_i) = (e^{\varphi - i\theta_i}, 1)^T/\sqrt{1+e^{+2\varphi}}$, with corresponding energy $E/m = +\cosh \varphi$. We can expand the wave function around the junction by the combination of incoming and outgoing edge states:
\begin{widetext}
	\begin{equation}
	\left\{
	\begin{aligned}
 \Psi_1(0) &= \bigg{(}\frac{e^{-\varphi - i\theta_1}}{\sqrt{1 + e^{-2\varphi}}} + r \frac{e^{\varphi - i\theta_1}}{\sqrt{1+e^{2\varphi}}} \bigg{)}\psi_{\alpha}(x_1 = 0) + \bigg{(} \frac{1}{\sqrt{1 + e^{-2\varphi}}} + \frac{r}{\sqrt{1+e^{2\varphi}}} \bigg{)} \psi_{\beta}(x_1=0), \\
 \Psi_2(0) &= t  \frac{e^{\varphi - i\theta_2}}{\sqrt{1+e^{2\varphi}}} \psi_\alpha(x_2 = 0) + t\frac{1}{\sqrt{1+e^{2\varphi}}} \psi_\beta(x_2= 0).
   \end{aligned}
   \right.
\end{equation}
\end{widetext}
As shown in Fig.[\ref{1DScattering}], $\psi_\alpha(x_1) = \psi_\beta(x_2)$ and $\psi_\beta(x_1) = \psi_\alpha(x_2)$ since they are from the same Chern insulator. For the SOTI Eq.[\ref{Bloch}], the wave function should be continuous at the junction:
\begin{equation}
\left\{
\begin{aligned}
	 \frac{e^{-\varphi - i\theta_1}}{\sqrt{1 + e^{-2\varphi}}} + r \frac{e^{\varphi - i\theta_1}}{\sqrt{1+e^{2\varphi}}} &= t\frac{1}{\sqrt{1+e^{2\varphi}}}, \\
	\frac{1}{\sqrt{1 + e^{-2\varphi}}} + r\frac{1}{\sqrt{1+e^{2\varphi}}} &= t  \frac{e^{\varphi - i\theta_2}}{\sqrt{1+e^{2\varphi}}}.
\end{aligned}
\right.
\end{equation}
By solving this we derive:
\begin{equation}\label{coScatter2}
\left\{
\begin{aligned}
	 r &=r^\prime = e^{\varphi}\frac{e^{i(\theta_1 + \theta_2)}-1}{e^{2\varphi}-e^{i(\theta_1 + \theta_2) }}, \\
	 t &= e^{i\theta_2}\frac{e^{2\varphi}-1}{e^{2\varphi}-e^{i(\theta_1 + \theta_2) }}, \\ 
	 t^\prime &= e^{i\theta_1}\frac{e^{2\varphi}-1}{e^{2\varphi}-e^{i(\theta_1 + \theta_2) }}. 
\end{aligned}
\right.
\end{equation}
The reflection and transmission coefficients $r$ and $t$ satisfy conservation of probability current:
\begin{equation}
	|r|^2 + |t|^2 = \frac{e^{2\varphi}(2-2\cos(\theta_1 + \theta_2)) + e^{4\varphi}+1-2e^{2\varphi}}{e^{4\varphi} +1 -2e^{2\varphi} \cos(\theta_1 + \theta_2)} = 1.
\end{equation}
Finally we have the scattering matrix for two terminal junction as:
\begin{equation}
	S = \begin{pmatrix}
		t & r \\
		r^\prime & t^\prime
	\end{pmatrix} = \frac{1}{e^{2\varphi}-e^{i(\theta_1 + \theta_2)}}
	\begin{pmatrix}
		\tilde t & \tilde r \\
		\tilde r^\prime & \tilde t^\prime
	\end{pmatrix} . 
\end{equation}
One can easily check that the scattering matrix is unitary $S^\dagger S = {\mathbf 1}$. The coefficients of scattering matrix, see in Eq.[\ref{coScatter2}] has simultaneous poles:
\begin{equation}
		e^{i\theta_1 + i\theta_2 } -e^{2\varphi} =0, ~2\varphi = i{(\theta_1 + \theta_2 + 2n\pi)},~ n \in \mathbb{Z},
\end{equation}
which stands for bound states localized at the junction with energy and fractional charge as:
\begin{equation}
	E = \cosh \varphi = \cos \bigg{(} \frac{\varphi}{2} \bigg{)},  ~N_s = -\frac{|\theta_2 + \theta_1|}{2\pi}.
\end{equation} 
Remember that $\theta_1$ here is equal to $-\theta_1$ in Sec. \ref{minimal} due to the flipping of $x_1$-leg's direction, the above results is in accordance with Eq.[\ref{1dSol}]. 
We further define $\eta$ as:
\begin{equation}
	\eta = \frac{r}{t} = \frac{(e^{i\theta_1}-e^{-i\theta_2})}{(e^{\varphi}-e^{-\varphi})}.
\end{equation}
The argument and the absolute value of $\eta$ are:
   \begin{equation}
   	\arg(\eta) = \arctan\bigg{(}\frac{\sin \theta_1 + \sin \theta_2}{\cos \theta_1 - \cos \theta_2}\bigg{)} =\frac{\pi}{2}-\frac{\theta_1-\theta_2}{2} ,
   \end{equation}
\begin{equation}
	|\eta|^2 = \frac{2 - 2\cos \theta_1 \cos \theta_2 + 2\sin \theta_1 \sin \theta_2}{e^{2\varphi} + e^{-2\varphi} - 2}  = \frac{\sin^2(\frac{\theta_1 + \theta_2}{2})}{\sinh^2 \varphi}.
\end{equation}
Thus the bound state energy can also be parametrized by reflection and transmission coefficents as:
\begin{equation}
	E^2 = \cosh^2 \varphi = \frac{|t|^2}{|r|^2}\sin^2(\frac{\theta_1+\theta_2}{2}) + 1.
\end{equation}

\subsection{Edge networks for a Y-junction}

In this section, we provide more details about the edge network description of a three-leg junction (``Y-junction'').

\subsubsection{Bound states from matching wave function}
Different from the scattering matrix approach in Sec. \ref{ScatteringMatrix}, here we get the same results by matching the trial wave function and validate that the poles of scattering states do correspond to localized states. We can derive the trial wave function by using the similar method in Sec. \ref{minimal}. Substitute the trial wave function $\chi{(x_i)}$ in to Eq.[\ref{EdgeHam}] for each leg independently, we find the modes localized at two ends of $i$-th edge, with $\chi{(x_i)} =e^{i\delta_i} (e^{i(\varphi - \theta_i)} ,1 )^T$ for energy $\epsilon_{i}^o = \cos \varphi$. This gives the relation between $a_i$ and $b_i$ on the same leg. More specifically: for the leg 1, we have $\chi{(x_1)} =e^{i\delta_1} (e^{i(\varphi - \theta_1)} ,1)^T$, with the basis $\Psi(x_1)=(\psi_2(x_1),\psi_3(x_1))^T$; for the leg 2, we have $\chi{(x_2)} =e^{i\delta_2} (e^{i(\varphi - \theta_2)}, 1)^T$, with the basis $\Psi(x_2)=(\psi_4(x_2),\psi_5(x_2))^T$; for the leg 3, we have $\chi{(x_3)} =e^{i\delta_3} (e^{i(\varphi - \theta_3)}, 1)^T$, with the basis $\Psi(x_3)=(\psi_6(x_3),\psi_1(x_3))^T$.

Due to the continuity of the bound state wave function, the boundary conditions are:
\begin{equation}
	\left\{
	\begin{aligned}
		&e^{i\delta_3}e^{i\alpha_1} \psi_1(x_3\rightarrow 0^+)  = e^{i\delta_1} e^{i(\varphi - \theta_1)}  \psi_{2}(x_1\rightarrow 0^+),\\
		&e^{i\delta_1}e^{i \alpha_2} \psi_3(x_1\rightarrow 0^+)  = e^{i\delta_2} e^{i(\varphi - \theta_2)} \psi_{4}(x_2\rightarrow 0^+),\\
		&e^{i\delta_2}e^{i\alpha_3} \psi_5(x_2\rightarrow 0^+) = e^{i\delta_3} e^{i(\varphi - \theta_3)}  \psi_{6}(x_3\rightarrow 0^+), 
	\end{aligned}
	\right.,
\end{equation}
where $\alpha_{i=1,2,3}$ are phase factors acquired across the junction as mentioned in main text. We also have $ \psi_1(x_3\rightarrow 0^+) = \psi_2(x_1\rightarrow 0^+), ~\psi_3(x_1\rightarrow 0^+) = \psi_4(x_2\rightarrow 0^+), ~\psi_5(x_2\rightarrow 0^+) = \psi_6(x_3\rightarrow 0^+)$ since they are the edge states from the same Chern insulator layer. With these we have:
\begin{equation}
	e^{i(\alpha_1 + \alpha_2 +  \alpha_3)} = e^{i(3\varphi - \theta_1 - \theta_2 - \theta_3 )},
\end{equation}
which is equivalent to
\begin{equation}\label{Energy}
\begin{aligned}
	&3\varphi = \sum_i (\theta_i + \alpha_i)  + 2n\pi,~ n \in {\mathbb Z}, \\
	&\frac{E}{m}= \cos \varphi, ~N_s = -\frac{\sum_i\theta_i}{2\pi}.
\end{aligned}
\end{equation}
This is in accordance with Eq.[\ref{ScatteringPole}] in main text. Thus the poles of the scattering matrix do correspond to the localized states at the junction. Similar results also apply for a vertex of octahedral or icosahedral topological fullerenes, as shown in Sec. \ref{ScatteringMatrix}.

\subsubsection{Y-junction scattering matrix}\label{AYScattering}
In this section we provide more details about how to derive the full scattering matrix Eq.[\ref{Matrix}] in Sec. \ref{ScatteringMatrix}. Similarly to the two terminal junction, for the scattering states of Y-junction, we denote $\eta = v/m$. 
We set $k\eta = \sinh \varphi >0$. Then for $x_i > 0$, under the basis $\big{(}\psi_\alpha(x_i),\psi_\beta(x_i)\big{)}^T$, for each individual leg, from Eq.[\ref{EdgeHam}] we can derive normalized wave function of incoming and outgoing modes as: $\psi_{\text{in}}^T(x_i) = (e^{-\varphi -i\theta_i}, 1)^T/\sqrt{1+e^{-2\varphi}},\quad \psi_{\text{out}}^T(x_i) = (e^{\varphi - i\theta_i}, 1)^T/\sqrt{1+e^{+2\varphi}}$, with corresponding energy $E/m = +\cosh \varphi$. As mentioned in main text, we first inject the mode along negative $x_1$ direction. The wave function on leg $x_1$ is given by $\Psi_1 = e^{-ikx_1}\psi_{1,\text{in}} + r e^{ikx_1} \psi_{1,\text{out}}$. Meanwhile, the wave function on leg $x_2$ is given by $\Psi_2 = t_{12} e^{ikx_2} \psi_{2,\text{out}}$, and the wave function on leg $x_3$ is given by $\Psi_3 = t_{13} e^{ikx_3} \psi_{3,\text{out}}$. We can expand the wave function around the intersection as:
\begin{widetext}
	\begin{equation}
		\left\{ \begin{aligned}
			\Psi_1(0) &= \bigg{(}\frac{e^{-\varphi - i\theta_1}}{\sqrt{1 + e^{-2\varphi}}} + r \frac{e^{\varphi - i\theta_1}}{\sqrt{1+e^{2\varphi}}} \bigg{)}\psi_{\alpha}(x_1 = 0) + \bigg{(} \frac{1}{\sqrt{1 + e^{-2\varphi}}} + \frac{r}{\sqrt{1+e^{2\varphi}}} \bigg{)} \psi_{\beta}(x_1=0), \\
            \Psi_2(0) &= t_{12}  \frac{e^{\varphi - i\theta_2}}{\sqrt{1+e^{2\varphi}}} \psi_\alpha(x_2 = 0) + t_{12}\frac{1}{\sqrt{1+e^{2\varphi}}} \psi_\beta(x_2= 0), \\
            \Psi_3(0) &= t_{13}  \frac{e^{\varphi - i\theta_3}}{\sqrt{1+e^{2\varphi}}} \psi_\alpha(x_3 = 0) + t_{13}\frac{1}{\sqrt{1+e^{2\varphi}}} \psi_\beta(x_3= 0).
		\end{aligned}
		\right.
	\end{equation}
\end{widetext}
Note that due to the continuity of edge state wave function for each individual layer, we have $\psi_\alpha(x_1 \rightarrow 0^+) = \psi_\beta(x_3 \rightarrow 0^+)$, $\psi_\beta(x_1 \rightarrow 0^+) = \psi_\alpha (x_2 \rightarrow 0^+)$, and $\psi_\beta(x_2 \rightarrow 0^+) = \psi_\alpha (x_3 \rightarrow 0^+)$. Following the assumption we made in Sec. \ref{Multi}, during the scattering process, the amplitude of the chiral edge states from same triangular Chern insulator is conserved, but they may acquire an additional phase factor $\alpha_i$ when by passing the junction. By matching the coeffients of $\psi_{\alpha(\beta),i}$ we have:
\begin{equation}
\left\{
\begin{aligned}
	& \frac{e^{-\varphi - i\theta_1}}{\sqrt{1 + e^{-2\varphi}}} + r \frac{e^{\varphi - i\theta_1}}{\sqrt{1+e^{2\varphi}}} = t_{13}\frac{1}{\sqrt{1+e^{2\varphi}}}e^{i\alpha_1}, \\
	&t_{12} \frac{e^{\varphi - i\theta_2}}{\sqrt{1+e^{2\varphi}}} =\bigg{(}\frac{1}{\sqrt{1 + e^{-2\varphi}}} + r\frac{1}{\sqrt{1+e^{2\varphi}}}\bigg{)} e^{i\alpha_2}, \\
	&t_{13} \frac{e^{\varphi-i\theta_3}}{\sqrt{1+ e^{2\varphi}}} = t_{12} \frac{1}{\sqrt{1+e^{2\varphi}}}e^{i\alpha_3}.
\end{aligned}
\right.
\end{equation}
From the above equation we derive that:
\begin{equation}
\left\{
\begin{aligned}
	r_1&= \frac{e^\varphi (e^{i\sum_i (\theta_i +\alpha_i)} -e^\varphi)}{e^{3\varphi} -e^{i\sum_i(\alpha_i+\theta_i)}},\\ 
	t_{12} &= \frac{e^{i(\alpha_2+\theta_2)}e^\varphi(e^{2\varphi}-1)}{e^{3\varphi} -e^{i\sum_i(\alpha_i+\theta_i)}}, \\
	 t_{13} &=\frac{e^{i(\alpha_2+\theta_2+\alpha_3 + \theta_3)}(e^{2\varphi}-1)}{e^{3\varphi} -e^{i\sum_i(\alpha_i+\theta_i)}}.
\end{aligned}
\right.
\end{equation}
One can check that the scattering is unitary:
\begin{equation}
	|r|^2 + |t_{12}|^2 + |t_{13}|^2 = 1.
\end{equation}

To derive the full scattering matrix, we can further inject the mode along negative $x_2$ ($x_3$) direction. By following the similar procedure for injecting along negative $x_1$ direction, we have:
\begin{equation}
\left\{
\begin{aligned}
	r_2 &= \frac{e^\varphi (e^{i\sum_i (\theta_i +\alpha_i)} -e^\varphi)}{e^{3\varphi} -e^{i\sum_i(\alpha_i+\theta_i)}},\\
	 t_{23} &= \frac{e^{i(\alpha_3+\theta_3)}e^\varphi(e^{2\varphi}-1)}{e^{3\varphi} -e^{i\sum_i(\alpha_i+\theta_i)}}, \\
	  t_{21} &=\frac{e^{i(\alpha_1+\theta_1+\alpha_3 + \theta_3)}(e^{2\varphi}-1)}{e^{3\varphi} -e^{i\sum_i(\alpha_i+\theta_i)}}.
\end{aligned}
\right.
\end{equation}
\begin{equation}
	\left\{
\begin{aligned}
	r_3& = \frac{e^\varphi (e^{i\sum_i (\theta_i +\alpha_i)} -e^\varphi)}{e^{3\varphi} -e^{i\sum_i(\alpha_i+\theta_i)}},\\
	 t_{31} &= \frac{e^{i(\alpha_1+\theta_1)}e^\varphi(e^{2\varphi}-1)}{e^{3\varphi} -e^{i\sum_i(\alpha_i+\theta_i)}}, \\
	  t_{32} &=\frac{e^{i(\alpha_1+\theta_1+\alpha_2 + \theta_2)}(e^{2\varphi}-1)}{e^{3\varphi} -e^{i\sum_i(\alpha_i+\theta_i)}}.
\end{aligned}
\right.
\end{equation}
Finally, we derive scattering matrix as:
\begin{equation}\label{SMatrix}
	S = \begin{pmatrix}
		r_1 & t_{12} & t_{13} \\
		t_{21} & r_2 & t_{23} \\
		t_{31} & t_{32} & r_3 
	\end{pmatrix} = \frac{1}{e^{3\varphi} - e^{i\Lambda}}\begin{pmatrix}
		\tilde r_1 & \tilde t_{12} & \tilde t_{13} \\
		\tilde t_{21} & \tilde r_2 & \tilde t_{23} \\
		\tilde t_{31} & \tilde t_{32} & \tilde r_3 
	\end{pmatrix},
	\end{equation}
where $k v/m = \sinh \varphi,~\Lambda= \sum_i (\theta_i+\alpha_i)$. The poles of $S$ denotes the existence of bound state with energy:
\begin{equation}
	 \frac{E}{m} = \cosh \varphi = \cos \bigg{(} \frac{\Lambda+2n\pi}{3} \bigg{)},~n\in {\mathbb Z},
\end{equation}
which is the Eq.[\ref{ScatteringPole}] in main text. It is easy to check that the Scattering matrix here is unitary, i.e., $S^\dagger S ={\mathbf 1}$. Eq.[\ref{ScatteringPole}] can be generalized to $l$-leg junction: ${E}/{m} = \cosh \varphi = \cos [{(\Lambda + 2n\pi)}/{l}], ~n \in \mathbb Z,~ N_s = - {\sum_i \theta_i}/{2\pi}$. For latter convenience we let $2n\pi$ be absorbed into $\sum_i\alpha_i$.

\subsubsection{Comparison with numerical results from exact diagonalization of tight-binding Hamiltonian}\label{ExactD}
\begin{figure}[h!]
\centering 
\includegraphics[width=1\columnwidth]{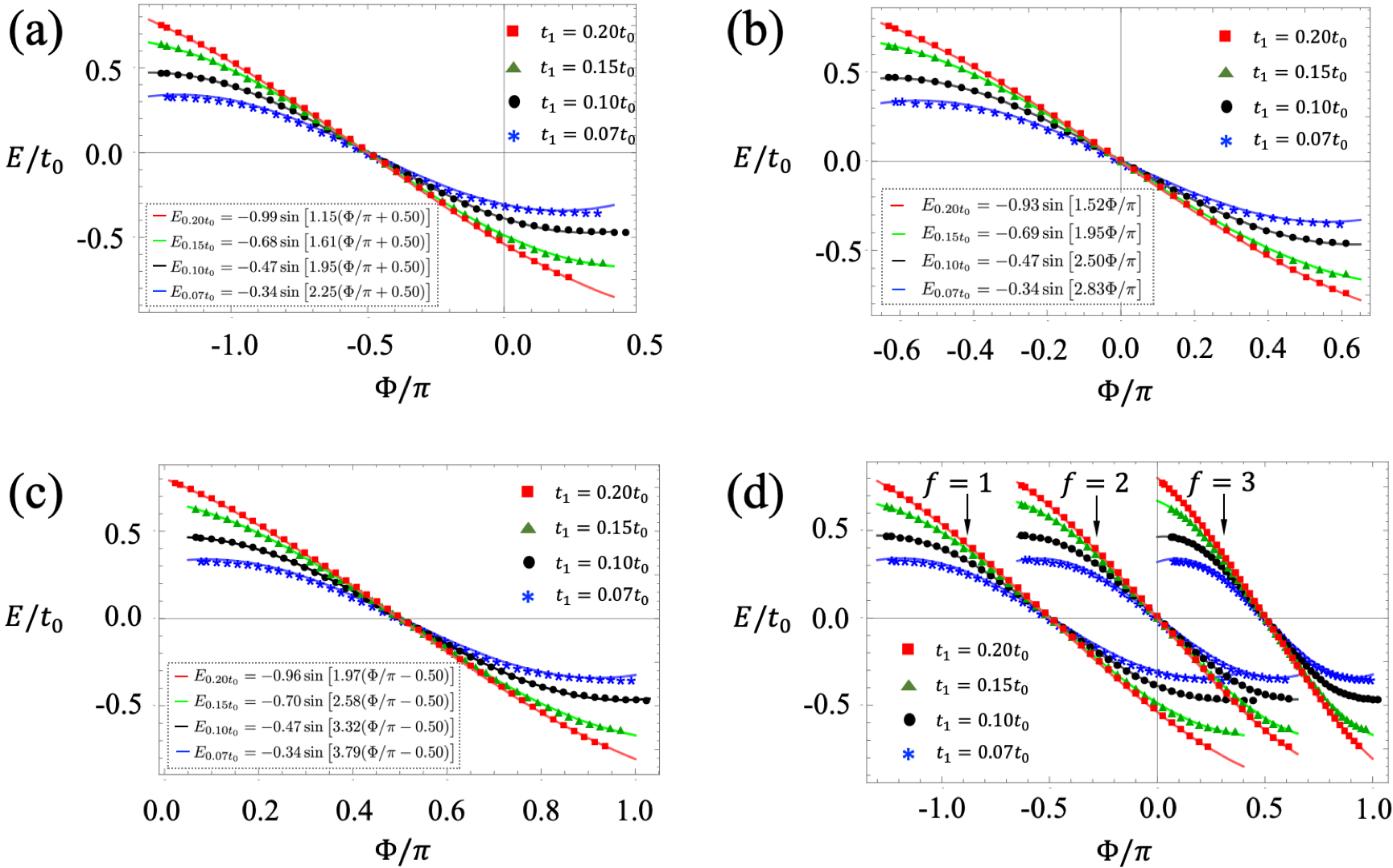}
\caption{\label{SuppTB} Bound state energy with external flux. The dots are from exactly diagonalizing the Haldane model Eq.[\ref{Haldane}]. The solid lines are fittings from exact diagonalization, which take the form of energy-phase(flux) relation Eq.[\ref{AnalyticEnergy}]. (a) Disclination with Frank index $f=1$. (b) Disclination with Frank index $f=2$. (c) Disclination with Frank index $f=3$. We also plot (a-c) in the same frame, as shown in (d). In (a-d), the red, green, black, and blue lines or dots denote, respectively, $t_1 = 2t_0$($m =1.04t_0$), $t_1 = 0.15t_0$($m=0.78 t_0$), $t_1 = 0.10 t_0$($m=0.52t_0$), and $t_1 = 0.07t_0$($m = 0.36t_0$). Here, $t_0$ stands for nearest neighbor hopping, $t_1$ stands for next-nearest neighbor hopping with $\phi_{ij}=\pi/2$., and $m=3\sqrt{3}t_1$ stands for Haldane mass, as shown in the main text. The equations on the left bottom side are the fitting of the numerical results from exact diagonalization. The calculation is done for 800 unit cells within each $60^\circ$ slice.}
\end{figure}
The bound-state energy Eq.[\ref{MultiLeg}] is depending on $\Lambda = \sum_i(\theta_i+\alpha_i)$. As we showed in main text, $\sum_i \alpha_i = A \sum_i \theta_i +B$, substitute these into Eq.[\ref{MultiLeg}] we have:
\begin{equation}\label{LinearEnergy}
	\frac{E}{m} = \cos\bigg{[}\frac{(1+A)\sum_i\theta_i+B }{6-f}\bigg{]},
\end{equation}
where $l = f-6$ is the number of legs for a disclination with Frank index $f$. In order to figure out the value of $A,B$ and derive the full response function as Eq.[\ref{Dis180}], in principle we need two data points (the bound state energy at two different flux value $\Phi$) from the exact diagonalizing tight-binding Hamiltonian Eq.[\ref{tbeqn}]. In fact, we do take two data points directly for $m\approx t_0$ and get Eq.[\ref{Dis180}] in the main text. However, note that for Frank index $f=3$ ($f=1$), although the response of bound state energy with respect to external flux for different $m$ are different, adding an external flux $\Phi = - \pi/2$ ($\Phi = + \pi/2$) can restore the particle hole symmetry, and move the bound state energy to zero. Thus we can define $\Phi_0=|{B}/({1+A})| = (\sum_i \theta_i)~\text{mod}~ \pi$, which is fixed for given $f$. Note that $\sum_i \theta_i$ is the total phase mismatch at the junction. With these Eq.[\ref{LinearEnergy}] can be reduced to:
\begin{equation}\label{AnalyticEnergy}
	\frac{E}{m} = \sin\bigg{[}\frac{(1+A)\pi(\Phi/\pi \pm \Phi_0/\pi)}{6-f}\bigg{]}.
\end{equation} 
The plus or minus sign here depends on whether the local Chern vector is align with or opposite to the direction of external flux. Now we only need one date point (for example, the energy of bound state in the absence of external flux) from exact diagonalization to get the value $A$ in Eq.[\ref{AnalyticEnergy}] and reproduce Eq.[\ref{Dis180}] directly. For $m = 3\sqrt{3}t_1 \approx t_0 =1$, we derive $A$ first and Eq.[\ref{AnalyticEnergy}] is then simplified as:
\begin{equation}\label{EnergyVsFlux}
\left\{
\begin{aligned}
	E_{\text{Tetrahedron}}(\Phi) &= \cos \bigg{(} \frac{2\Phi}{3}+ \frac{\pi}{6}\bigg{)}, \\
	E_{\text{Octahedron}}(\Phi) &= \cos \bigg{(} \frac{2\Phi}{4}+ \frac{\pi}{2}\bigg{)}, \\
	E_{\text{Icosahedron}}(\Phi) &= \cos \bigg{(} \frac{2\Phi}{5}+ \frac{7\pi}{10}\bigg{)},
\end{aligned}
\right.
\end{equation}
which is Eq.[\ref{Dis180}] in the presence of external flux $\Phi$.

We further compare the results from Eq.[\ref{Dis180}] with full numerical results derived from exactly diagonalizing the tight-binding Hamiltonian Eq.[\ref{Haldane}], as shown in Fig.[\ref{SuppTB}]. We plot the bound state energy with external flux ($\Phi$) under different Haldane mass $m = 3\sqrt{3}t_1$ and different Frank index $f$. The direct fittings of numerical results do take the form of Eq.[\ref{AnalyticEnergy}], as shown in the left-bottom of each sub-figure.

\subsubsection{Comparison to numerical results from continuous model}
\begin{figure}[!h]
\centering 
\includegraphics[width=1\columnwidth]{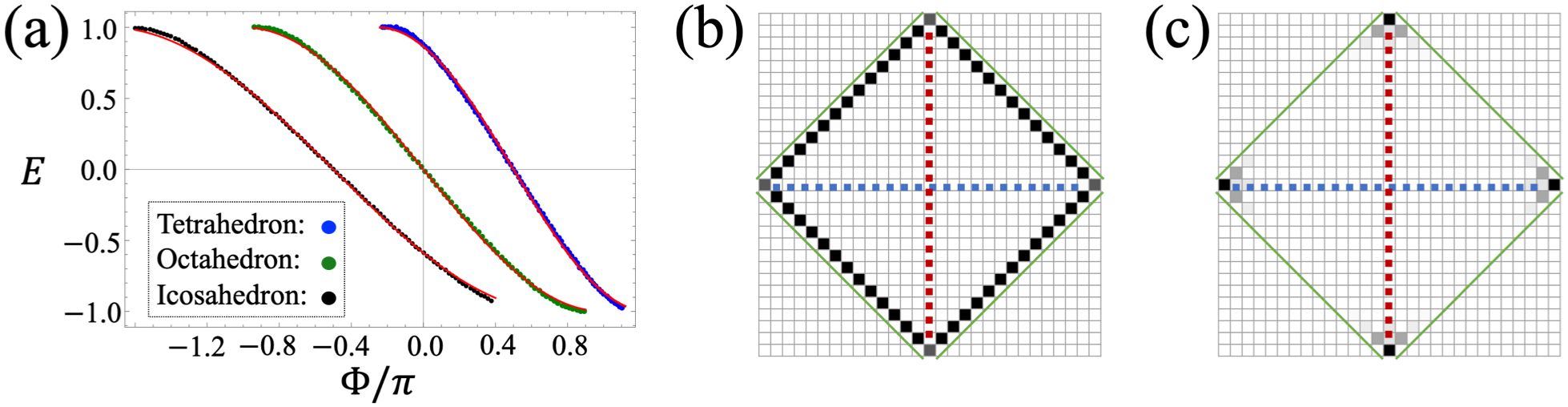}
\caption{\label{SuppFlux}(a) Bound state energy with external flux. The dots are numerical results for solving Eq.[\ref{BoundEnergy}]. The blue dots are for one vertex of tetrahedral topological fullerenes (disclination with Frank index $f=3$). The green dots are for one vertex of octahedral topological fullerenes (disclination with Frank index $f=2$). The black dots are for one vertex of icosahedral topological fullerenes (disclination with Frank index $f=1$). The red lines are relevant results from Eq.[\ref{Energy}]. (b-c) The wave function density for mid gap state in (b) Quadrupole insulator, and (c) 2d SOTI from Eq.[\ref{tbeqn}] proposed in main text. The red and blue dashed lines stand for the reflection symmetric axes for $x-$ and $y-$ directions, respectively. The green solid lines stand for the boundary.}
\end{figure}

The bound state energy with respect to external flux from continuous model for conical singularities~\cite{Ruegg2013prl} is given by:
\begin{equation}\label{BoundEnergy}
	\sqrt{\frac{m-E}{m+E}} = \frac{K_{\nu-1/2}(\kappa \rho)}{K_{\nu+1/2}(\kappa \rho)}
\end{equation}
where
\begin{equation}
	 \kappa = \sqrt{m^2 - E^2}, ~ \nu = \frac{j - \frac{\Phi}{2\pi} + \frac{f}{4}}{1-\frac{f}{6}}.
\end{equation}
Here $m$ is the Haldane mass, $\rho$ stands for the radius of the hole in disclination, $E$ is the bound state energy, $j$ is half integer, $f$ is Frank index and stands for the number of $\pi/3$ wedges removed, $\Phi$ denotes the external flux, and $K(\kappa\rho)$ is modified Bessel functions of the second kind. We have set the positive direction of external flux opposite to local Chern vector. In practice, in order to derive full energy-flux relation for given $m$, one may need (at least) one data point (bound state energy at given $\Phi$) from exact diagonalizing Eq.[\ref{tbeqn}] to get the value of effective radius $\rho$. After that we can derive the bound state energy with external flux from (numerically) solving Eq.[\ref{BoundEnergy}].

We have shown that our analytic results in Eq.[\ref{AnalyticEnergy}] fit quite well with the numerical results from diagonalizing tight-binding model in previous subsection. Our method also give the proper results from solving Eq.[\ref{BoundEnergy}] directly, as shown in Fig.[\ref{SuppFlux}.(a)]. From here we know that, the phase shift $\sum_i \alpha_i$ should be a function of Haldane mass $m$ and effective radius $\rho$.

\subsection{Boundary Hamiltonian for arbitrary edge}
In this section we derive the effective edge Hamiltonian for an arbitrary edge. Note that the in gap state wave function distribution for our Tetrahedral type TI (Eq.[\ref{Bloch}]) is different from that of Quadrupole insulator, see in Fig.[\ref{SuppFlux}.(b,c)]. We further show that our Tetrahedral type 2d SOTI can hold fractional charge at the corner of rectangular boundaries, regardless of the orientation of the rectangle.

\begin{figure}[!h]
\centering 
\includegraphics[width=1\columnwidth]{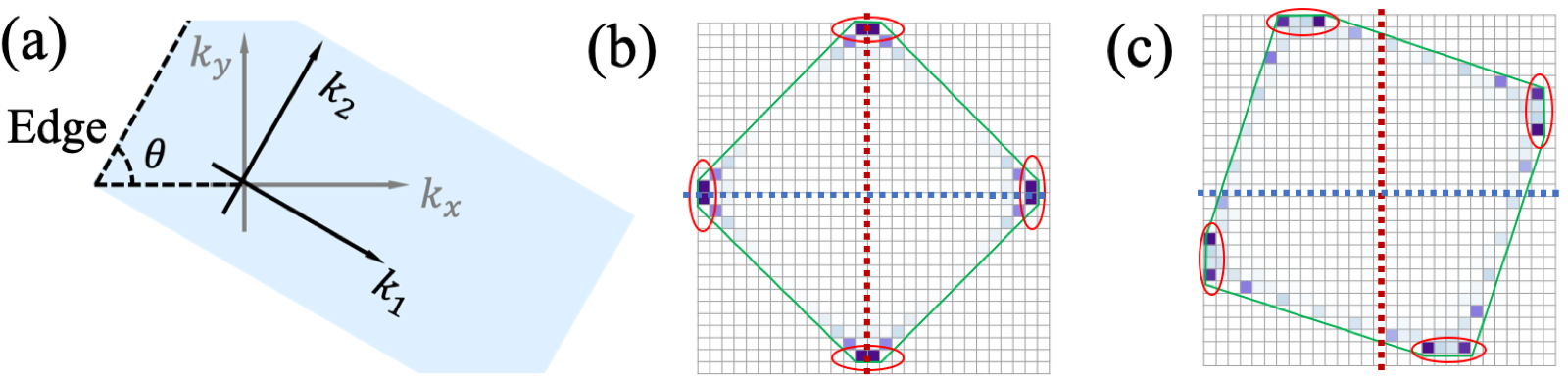}
\caption{\label{ArbitraryEdge}(a) Edge along $\hat e_2 = \cos \theta \hat e_x + \sin \theta  \hat e_y$ direction (marked by dashed line). (b,c) Corner charge (in gap state wave function density) in the presence of different boundary conditions. The dashed blue and red lines stand for two reflection symmetric axes, the green solid line denotes the boundary of tetrahedral type TI. The corner localized charge is marked by red circles. (b) Boundary configuration respects original reflection symmetry. (c) Boundary configuration does not respect original reflection symmetry.}
\end{figure}

The Bloch Hamiltonian for our Tetrahedral type TI is Eq.[\ref{Bloch}], as shown in main text. In the absence of inter-layer coupling, i.e. $\lambda =0$, the system can be viewed as index spin hall effect:
\begin{equation}\label{QSHEq}
\begin{aligned}
	{\mathcal H}(\vec k)_{\text{QSH}}= &2 t_{\text{so}} \sin{(k_x)}\tau_1  +2t_{\text{so}} \sin{(k_y)} \tau_2 \\
	+&(m_z - 2t_x \cos{k_x} - 2t_y\cos{k_y}) \tau_3 \sigma_3.
\end{aligned}
\end{equation}
Around $(k_x = 0, k_y = 0)$, the low energy version for Hamiltonian Eq.[\ref{QSHEq}] is given by:
\begin{equation}\label{lowEne}
	h(\vec k) = 2t_{\text{so}} k_x \tau_1 + 2t_{\text{so}} k_y \tau_2 + (\tilde m_z + t_x k_x^2 + t_y k_y^2 ) \tau_3 \sigma_3, 
\end{equation}
where $\tilde m_z = m_z -2t_x - 2t_y$. For simplicity we assume $t_x = t_y = t_0$.

In order to figure out the edge states at the cut along $\vec e_2 = \cos \theta \hat e_x + \sin \theta  \hat e_y$ direction (see in Fig.[\ref{ArbitraryEdge}].(a)), we define a new set of basis in both spatial and momentum spaces:
\begin{equation}\label{basis}
\left\{
\begin{aligned}
	x& = x_1 \sin \theta + x_2 \cos \theta, \\
	 y &= -x_1 \cos \theta + x_2 \sin \theta, 
\end{aligned}
\right. ~ \left\{
\begin{aligned}
	k_x &= k_1 \sin \theta + k_2 \cos \theta, \\ k_y &= -k_1 \cos \theta + k_2 \sin \theta.
\end{aligned}
\right.
\end{equation}
Substituting Eq.[\ref{basis}] into Eq.[\ref{lowEne}], the Low energy Hamiltonian can be written in the form of $k_{1,2}$:
\begin{equation}\label{reLowEne}
\begin{aligned}
	h(\vec k) =& 2t_{\text{so}} (k_1 \sin \theta + k_2 \cos \theta)\tau_1 + (\tilde m_z + t_0 k_1^2 + t_0 k_2^2 ) \tau_3 \sigma_3\\
	+&2t_{\text{so}} (-k_1 \cos \theta + k_2 \sin \theta) \tau_2 .
\end{aligned}
\end{equation}
Consider the model Hamiltonian Eq.[\ref{reLowEne}] defined on the half-space $x_1 >0$ in the $x_1-x_2$ plane. We replace $k_1 \rightarrow -i\partial_{x_1}, k_2 \rightarrow 0$, and neglect the higher order terms in Eq.[\ref{reLowEne}]:
\begin{equation}
\begin{aligned}
	\tilde h(x_1) = (-i\partial_{x_1} 2t_{\text{so}}\sin \theta) \sigma_1 + (i\partial_{x_1} 2t_{\text{so}}\cos \theta) \sigma_2 + \tilde m_z \sigma_3 \tau_3.
\end{aligned}
\end{equation}
By using the ansatz $\psi_0 = e^{\eta x_1}\phi$, we can find a pair of counter-propagating edge states:
\begin{equation}
\left\{
\begin{aligned}
	\Psi_\uparrow &= \frac{e^{-2t_{\text{so}}x_1/\tilde m_z}}{\sqrt{N_\uparrow}} (\cos \theta - i \sin \theta,1,0,0)^{T}, \\ \Psi_\downarrow &= \frac{e^{-2t_{\text{so}}x_1/\tilde m_z}}{\sqrt{N_\downarrow}} (0,0,-\cos \theta + i \sin \theta,1)^{T},
\end{aligned}
\right.
\end{equation}
where $N_{\uparrow(\downarrow)}$ is the normalization constant.
This procedure~\cite{Qi2011} leads to a $2\times 2$ effective Hamiltonian defined by $H^{\alpha,\beta}_{\text{edge}}(k_2) = \bra{\Psi_\alpha} h(\vec k) \ket{\Psi_\beta}$, to the leading order in $k_2$, we arrive at the effective Hamiltonian for helical edge states:
\begin{equation}
	h_{\text{edge}}^0 = 2t_{so} k_2 \sigma_z.
\end{equation}
Similarly, the inter-layer coupling $\lambda \tau_1 \sigma_1$ (or $\lambda \tau_2 \sigma_2$), under the basis $\Psi_{\alpha,\beta}$, gives birth to an additional term $-\lambda \sin \theta \sigma_y$ (or $- \lambda \cos \theta \sigma_x$). In summary, under the basis $\Psi_{\alpha,\beta}$, the total effective edge Hamiltonian is given by:
\begin{equation}\label{lowenergyApp}
	h_{\text{edge}} = 2t_{so} k_2 \sigma_z  -\lambda \cos \theta \sigma_x - \lambda \sin \theta \sigma_y.
\end{equation}
We can define the effective mass term as:
\begin{equation}
\begin{aligned}
	{\mathcal M}_i &= - \lambda( \cos \theta_i \sigma_x +  \sin \theta_i \sigma_y) \\
	&=- \lambda (\cos \theta_i \hat e_x + \sin \theta_i \hat e_y) \cdot{} (\sigma_x \hat e_x + \sigma_y \hat e_y + \sigma_z \hat e_z) \\
	&= -\lambda \vec e_i \cdot{} \vec \sigma.
\end{aligned}
\end{equation}
This related the effective mass term of $i$-th edge to its orientation $\hat e_i = \cos \theta_i \vec e_x + \sin \theta_i \vec e_y$. According to our previous results, the kink of effective mass term at the corner can give birth to corner localized charge. The value of the charge (edge soliton) $N_s$ is:
\begin{equation}
	N_s = -\frac{\theta_2 - \theta_1}{2} = -\frac{\delta \theta}{2}.
\end{equation}
For any rectangular boundary, $\delta \theta = \pi/2$ since two adjoint edges are perpendicular to each other. Thus the corner localized fractional charge should be $-1/4e$, regardless the orientation of rectangle. We have confirmed this by exact diagonalizing the tight-binding Hamiltonian, as shown in Fig.[\ref{ArbitraryEdge}.(b,c)]. This result can be generalized to the corner state with arbitrary fractional charge by tuning the angle $\theta$ between two adjoint edges.

\subsection{Fractional charge for edge soliton}
In the absence of particle hole symmetry, the domain wall state for a SSH chain can hold bound state with non-zero energy and fractional charge aside from $-1/2e$~\cite{Jackiw1983, Goldstone1981}. In this section, we summarized and slightly modified their previous works~\cite{Jackiw1983} and derive the similar results for edge solitons. This is in accordance with the results Eq.[\ref{TopoC}] from bosonization in Sec. \ref{Description}.

Suppose we have a one-dimensional Dirac Hamiltonian in the external field $\varphi$:
\begin{equation}\label{Dirac}
	\hat H(\varphi) = -i\partial_x \sigma_z + \epsilon \sigma_x + \varphi(x) \sigma_y.
\end{equation}
For simplicity we assume that $\epsilon > 0$. Up to a global normalization constant and a unitary transformation this Hamiltonian can be connected to Hamiltonian Eq.[\ref{lowenergyApp}]. In the absence of $\epsilon \sigma_x$, the Hamiltonian respects the charge conjugation symmetry and can hold zero mode when $\varphi(x)$ has a kink. The presence of $\epsilon \sigma_x$ breaks the charge conjugation symmetry of the system. We will see later on that this Hamiltonian can hold bound state with nonzero energy and fractional charge.

In the vacuum where the system does not hold a soliton, $\varphi = \varphi_0 = \text{const}$. We denote $\varphi_0 = \mu$ for simplicity. In the presence of a soliton, $\varphi(x) = \varphi_s(x)$, and in principle the $\varphi_s(x)$ should have a kink. In order to compute the charge, we need to derive the eigenstates of this two situation. The Schrodinger equation for these two scenarios can be written as:
\begin{equation}\label{SEQ}
	\hat H(\varphi_0) \psi_E^0 = E^0 \psi_E^0, ~ \hat H(\varphi_s) \psi^s_E =E^s \psi_E^s,
\end{equation}
where $\psi_E^0$ stands for the normal state without solitons, $\psi^s_E$ stand for the situation in the presence of soliton.

The charge density at level $E$ is $\rho_E(x) = \psi_E^\dagger(x) \times \psi_E(x)$, and the physical charge density is got by integrating $\rho_E$ over all negative $E$, since the negative energy levels are filled in the half-filling:
\begin{equation}
	\rho(x) = \int_{-\infty}^0 dE \rho_E(x).
\end{equation}
Finally the soliton charge is obtained by integrating the charge density in the soliton field over all $x$, but to avoid an infinity, we must subtract a similar integral of the charge density when no soliton is present:
\begin{equation}
	Q = \int dx (\rho_s(x) - \rho_0(x)).
\end{equation}
We can calculate the exact value of $Q$ even if we do not know the exact form of $\varphi(x)$. All we need to know about $\varphi_s$ is that it interpolates between opposite ``vacuum'' values as $x$ passes from $-\infty $ to $+\infty$:
\begin{equation}
	\varphi_s(+\infty) = |\varphi_0|=\mu,~ \varphi_s(-\infty)= - |\varphi_0| = -\mu.
\end{equation}
We now study the eigenstates of Eq.[\ref{SEQ}]. The vacuum problem is trivial: the wave functions are plane waves $\propto e^{ikx}$ and the spectrum is continuous $E^0 = \pm (k^2 + \mu^2 + \epsilon^2)^{1/2}$.

In the presence of soliton, we first assume that the wave-function of the eigenstate is $(u,v)^T$. Thus we have:
\begin{equation}
	\begin{pmatrix}
		-i\partial_x & -i\varphi(x)+\epsilon \\
		i\varphi(x) +\epsilon & i\partial_x
	\end{pmatrix}
	\begin{pmatrix}
		u \\
		v
	\end{pmatrix}
	=E \begin{pmatrix}
		u \\ v
	\end{pmatrix},
\end{equation}
which can be simplified as:
\begin{equation}
\left\{
	\begin{aligned}
		&-i\partial_x u + (-i\varphi(x) + \epsilon) v = E u, \\
		&(i\varphi(x) + \epsilon) u + i\partial_x v = E v.
	\end{aligned}
\right.
\end{equation}
In order to solve these two equations, we first add up two equations:
\begin{equation}
	-i\partial_x(u-v)  + i \varphi(x) (u - v) + \epsilon(u + v) = E(u+v),
\end{equation}
and then subtract the second equation from the first one, such that we have:
\begin{equation}
	-i\partial_x(u+v) - i\varphi(x)(u + v) - \epsilon(u-v) = E(u - v).
\end{equation}
We define the new parameters:
\begin{equation}
	U = \frac{u +v}{\sqrt{2}}, \quad V = \frac{u -v}{\sqrt{2}},
\end{equation}
then we can rewrite the result as:
\begin{equation}\label{Dlinear}
\left\{
	\begin{aligned}
		&(-i\partial_x + i\varphi(x) )V = (E-\epsilon) U, \\
		&(-i\partial_x - i\varphi(x)) U = (E+\epsilon) V.
	\end{aligned}
\right.
\end{equation}
From the second line of Eq.[\ref{Dlinear}] we know that 
\begin{equation}
	V = \frac{-i(\partial_x + \varphi(x))}{E+\epsilon}  U.
\end{equation}
Substitute this into the first line of Eq.[\ref{Dlinear}], we have:
\begin{equation}\label{second}
	-(\partial_x^2 - \varphi^2(x) + \partial_x \varphi(x)) U = (E^2 - \epsilon^2) U.
\end{equation}
From Eq.[\ref{Dlinear}] and Eq.[\ref{second}] we can figure out a possible solution: 
\begin{equation}
	U = \exp [-\int^x dx^\prime \varphi_x(x^\prime)], \quad V = 0,
\end{equation}
corresponds to the energy $E = \epsilon$. Note that the $U$ is localized at the kink $x = 0$ due to the form of $\varphi_s(x)$. 

To calculate the particle density, we still need to know the eigenstate for all negative energy solutions. We assume that $U \propto e^{ikx}$ and $\varphi \approx \pm \mu$ at large $x$ limit, thus we have the normalized factor:
\begin{equation}
	1 = |u_k|^2 + |v_k|^2 = |U_k|^2 + |V_k|^2  = U^2\frac{2E}{E+\epsilon},
\end{equation}
from which we can figure out the normalized wave function for the negative energy:
\begin{equation}
	U = \sqrt{\frac{E+\epsilon}{2E}} U_k, \quad V= -\frac{i}{\sqrt{2E(E+\epsilon)}}(\partial_x + \varphi(x)) U_k.
\end{equation}
This gives the wave function in originally basis:
\begin{equation}
	\psi_k = \begin{pmatrix}
		u_k \\ v_k
	\end{pmatrix},
\end{equation}
where
\begin{equation}
\left\{
\begin{aligned}
	u_k &= \frac{1}{\sqrt{2}} \bigg{(} \sqrt{\frac{E+\epsilon}{2E}} -\frac{i}{\sqrt{2E(E+\epsilon)}}(\partial_x + \varphi(x))  \bigg{)} U_k, \\
	v_k &= \frac{1}{\sqrt{2}} \bigg{(} \sqrt{\frac{E+\epsilon}{2E}} +\frac{i}{\sqrt{2E(E+\epsilon)}}(\partial_x + \varphi(x)) \bigg{)} U_k.
\end{aligned}
\right.
\end{equation}
The wave function $\psi_k$ satisfies: 
\begin{equation}
	\hat H (\varphi) \psi_k = E\psi_k, \quad E= -\sqrt{k^2 + \mu^2 + \epsilon^2}.
\end{equation}
The Charge-density at negative $E$ is given by:
\begin{equation}\label{density}
\begin{aligned}
	\rho_k(x) &= |u_k|^2 + |v_k|^2 \\
	&=[(E+\epsilon/2E)]|U_k|^2 + [2E(E+\epsilon)]^{-1}|(\partial_x + \varphi)U_k(x)|^2 \\
	&= |U_k(x)|^2 + [4E(E+\epsilon)]^{-1}\partial_x^2 |U_k(x)|^2 \\
	&~+ [2E(E+\epsilon)]^{-1}\partial_x[|U_k(x)|^2 \varphi(x)]
\end{aligned}
\end{equation}
where the validity of second line comes from Eq.[\ref{second}]. 

The soliton charge is the integral over all $x$ and $k$ above evaluated with $\varphi = \varphi_s$, minus a similar integral in the vacuum; but in the vacuum, $|U_k|^2 \equiv \varphi(x) = \mu$, such that the last two term in Eq.[\ref{density}] vanished. Thus we have the soliton charge:
\begin{widetext}
\begin{equation}\label{charge}
	N_s = \int dx \int_{-\infty}^{+\infty} \frac{dk}{2\pi} [|U_k^s(x)|^2 - |U_k^0(x)|^2] + \int_{-\infty}^{\infty} \frac {dk}{2\pi} \frac{1}{4E(E+\epsilon)}[\partial_x |U_k^s(x)|^2 + 2|U_k^s(x)|^2 \varphi_s(x)]|^{x=+\infty}_{x=-\infty}.
\end{equation}
\end{widetext}
The double integral can be evaluated by completeness: The $U^0_k$ represent all the Schrodinger modes in the vacuum, while the $U_k^s$ are one short of being complete in the soliton sector, since the normalized bound state is not among them. Hence the first term contributes $-1$ to $Q$. To evaluate the second term in Eq.[\ref{charge}], let us consider the wave function in the presence of a soliton when $x = \pm \infty$. These may be given in terms of transmission ($T$) and reflection coefficients ($R$):
\begin{equation}
	U_k^s( + \infty)  = T e^{ikx}, \quad U_k^s(-\infty)= e^{ikx} + R e^{-ikx}. 
\end{equation}
Thus, upon dropping oscillatory terms, we are left with the soliton charge:
\begin{equation}
	N_s = -1 + \int_{-\infty}^{+\infty} \frac{dk}{2\pi} \frac{\mu}{2E(E+\epsilon)}[|T|^2 + (|R|^2 + 1)],
\end{equation}
where the plus sign between the contributions at $x = + \infty$ and at $x = -\infty$ arises because of sign reversal in $\varphi_s(x)$. Unitarity, $|T|^2 + |R|^2 = 1$, permits a final evaluation:
\begin{equation}\label{Nscharge}
	N_s= - \frac{1}{\pi} \arctan \bigg{(}\frac{\mu}{\epsilon}\bigg{)}.
\end{equation}
Note that, if we denote $\epsilon = m \cos \theta$ and $\varphi_s(\pm \infty) = \pm \mu = m\sin (\mp \theta)$, Eq.[\ref{Nscharge}] is reduced to:
\begin{equation}
	N_s = -\frac{1}{\pi} \arctan (\tan\theta) = - \frac{\theta - (-\theta)}{2\pi},
\end{equation}
which is in accordance with Eq.[\ref{TopoC}] derived from bosonization of the helical Luttinger liquid.



\bibliography{edge_network_v3}

\begin{thebibliography}{56}%
\makeatletter
\providecommand \@ifxundefined [1]{%
 \@ifx{#1\undefined}
}%
\providecommand \@ifnum [1]{%
 \ifnum #1\expandafter \@firstoftwo
 \else \expandafter \@secondoftwo
 \fi
}%
\providecommand \@ifx [1]{%
 \ifx #1\expandafter \@firstoftwo
 \else \expandafter \@secondoftwo
 \fi
}%
\providecommand \natexlab [1]{#1}%
\providecommand \enquote  [1]{``#1''}%
\providecommand \bibnamefont  [1]{#1}%
\providecommand \bibfnamefont [1]{#1}%
\providecommand \citenamefont [1]{#1}%
\providecommand \href@noop [0]{\@secondoftwo}%
\providecommand \href [0]{\begingroup \@sanitize@url \@href}%
\providecommand \@href[1]{\@@startlink{#1}\@@href}%
\providecommand \@@href[1]{\endgroup#1\@@endlink}%
\providecommand \@sanitize@url [0]{\catcode `\\12\catcode `\$12\catcode
  `\&12\catcode `\#12\catcode `\^12\catcode `\_12\catcode `\%12\relax}%
\providecommand \@@startlink[1]{}%
\providecommand \@@endlink[0]{}%
\providecommand \url  [0]{\begingroup\@sanitize@url \@url }%
\providecommand \@url [1]{\endgroup\@href {#1}{\urlprefix }}%
\providecommand \urlprefix  [0]{URL }%
\providecommand \Eprint [0]{\href }%
\providecommand \doibase [0]{http://dx.doi.org/}%
\providecommand \selectlanguage [0]{\@gobble}%
\providecommand \bibinfo  [0]{\@secondoftwo}%
\providecommand \bibfield  [0]{\@secondoftwo}%
\providecommand \translation [1]{[#1]}%
\providecommand \BibitemOpen [0]{}%
\providecommand \bibitemStop [0]{}%
\providecommand \bibitemNoStop [0]{.\EOS\space}%
\providecommand \EOS [0]{\spacefactor3000\relax}%
\providecommand \BibitemShut  [1]{\csname bibitem#1\endcsname}%
\let\auto@bib@innerbib\@empty
\bibitem [{\citenamefont {Kane}\ and\ \citenamefont {Mele}(2005)}]{Kane2005}%
  \BibitemOpen
  \bibfield  {author} {\bibinfo {author} {\bibfnamefont {C.~L.}\ \bibnamefont
  {Kane}}\ and\ \bibinfo {author} {\bibfnamefont {E.~J.}\ \bibnamefont
  {Mele}},\ }\href {\doibase 10.1103/PhysRevLett.95.146802} {\bibfield
  {journal} {\bibinfo  {journal} {Phys. Rev. Lett.}\ }\textbf {\bibinfo
  {volume} {95}},\ \bibinfo {pages} {146802} (\bibinfo {year}
  {2005})}\BibitemShut {NoStop}%
\bibitem [{\citenamefont {Fu}\ \emph {et~al.}(2007)\citenamefont {Fu},
  \citenamefont {Kane},\ and\ \citenamefont {Mele}}]{Fu2007Mar}%
  \BibitemOpen
  \bibfield  {author} {\bibinfo {author} {\bibfnamefont {L.}~\bibnamefont
  {Fu}}, \bibinfo {author} {\bibfnamefont {C.~L.}\ \bibnamefont {Kane}}, \ and\
  \bibinfo {author} {\bibfnamefont {E.~J.}\ \bibnamefont {Mele}},\ }\href
  {\doibase 10.1103/PhysRevLett.98.106803} {\bibfield  {journal} {\bibinfo
  {journal} {Phys. Rev. Lett.}\ }\textbf {\bibinfo {volume} {98}},\ \bibinfo
  {pages} {106803} (\bibinfo {year} {2007})}\BibitemShut {NoStop}%
\bibitem [{\citenamefont {Moore}\ and\ \citenamefont
  {Balents}(2007)}]{moore&balents-2006}%
  \BibitemOpen
  \bibfield  {author} {\bibinfo {author} {\bibfnamefont {J.~E.}\ \bibnamefont
  {Moore}}\ and\ \bibinfo {author} {\bibfnamefont {L.}~\bibnamefont
  {Balents}},\ }\href {\doibase 10.1103/PhysRevB.75.121306} {\bibfield
  {journal} {\bibinfo  {journal} {Phys. Rev. B}\ }\textbf {\bibinfo {volume}
  {75}},\ \bibinfo {pages} {121306} (\bibinfo {year} {2007})}\BibitemShut
  {NoStop}%
\bibitem [{\citenamefont {Fu}\ and\ \citenamefont {Kane}(2007)}]{Fu2007July}%
  \BibitemOpen
  \bibfield  {author} {\bibinfo {author} {\bibfnamefont {L.}~\bibnamefont
  {Fu}}\ and\ \bibinfo {author} {\bibfnamefont {C.~L.}\ \bibnamefont {Kane}},\
  }\href {\doibase 10.1103/PhysRevB.76.045302} {\bibfield  {journal} {\bibinfo
  {journal} {Phys. Rev. B}\ }\textbf {\bibinfo {volume} {76}},\ \bibinfo
  {pages} {045302} (\bibinfo {year} {2007})}\BibitemShut {NoStop}%
\bibitem [{\citenamefont {Qi}\ and\ \citenamefont {Zhang}(2011)}]{Qi2011}%
  \BibitemOpen
  \bibfield  {author} {\bibinfo {author} {\bibfnamefont {X.-L.}\ \bibnamefont
  {Qi}}\ and\ \bibinfo {author} {\bibfnamefont {S.-C.}\ \bibnamefont {Zhang}},\
  }\href {\doibase 10.1103/RevModPhys.83.1057} {\bibfield  {journal} {\bibinfo
  {journal} {Rev. Mod. Phys.}\ }\textbf {\bibinfo {volume} {83}},\ \bibinfo
  {pages} {1057} (\bibinfo {year} {2011})}\BibitemShut {NoStop}%
\bibitem [{\citenamefont {Hasan}\ and\ \citenamefont {Kane}(2010)}]{Hasan2010}%
  \BibitemOpen
  \bibfield  {author} {\bibinfo {author} {\bibfnamefont {M.~Z.}\ \bibnamefont
  {Hasan}}\ and\ \bibinfo {author} {\bibfnamefont {C.~L.}\ \bibnamefont
  {Kane}},\ }\href {\doibase 10.1103/RevModPhys.82.3045} {\bibfield  {journal}
  {\bibinfo  {journal} {Rev. Mod. Phys.}\ }\textbf {\bibinfo {volume} {82}},\
  \bibinfo {pages} {3045} (\bibinfo {year} {2010})}\BibitemShut {NoStop}%
\bibitem [{\citenamefont {Moore}(2010)}]{Moore2010}%
  \BibitemOpen
  \bibfield  {author} {\bibinfo {author} {\bibfnamefont {J.~E.}\ \bibnamefont
  {Moore}},\ }\href {\doibase 10.1038/nature08916} {\bibfield  {journal}
  {\bibinfo  {journal} {Nature}\ }\textbf {\bibinfo {volume} {464}},\ \bibinfo
  {pages} {194 } (\bibinfo {year} {2010})}\BibitemShut {NoStop}%
\bibitem [{\citenamefont {Benalcazar}\ \emph
  {et~al.}(2017{\natexlab{a}})\citenamefont {Benalcazar}, \citenamefont
  {Bernevig},\ and\ \citenamefont {Hughes}}]{Benalcazar61}%
  \BibitemOpen
  \bibfield  {author} {\bibinfo {author} {\bibfnamefont {W.~A.}\ \bibnamefont
  {Benalcazar}}, \bibinfo {author} {\bibfnamefont {B.~A.}\ \bibnamefont
  {Bernevig}}, \ and\ \bibinfo {author} {\bibfnamefont {T.~L.}\ \bibnamefont
  {Hughes}},\ }\href {\doibase 10.1126/science.aah6442} {\bibfield  {journal}
  {\bibinfo  {journal} {Science}\ }\textbf {\bibinfo {volume} {357}},\ \bibinfo
  {pages} {61} (\bibinfo {year} {2017}{\natexlab{a}})}\BibitemShut {NoStop}%
\bibitem [{\citenamefont {Benalcazar}\ \emph
  {et~al.}(2017{\natexlab{b}})\citenamefont {Benalcazar}, \citenamefont
  {Bernevig},\ and\ \citenamefont {Hughes}}]{Benalcazar2017}%
  \BibitemOpen
  \bibfield  {author} {\bibinfo {author} {\bibfnamefont {W.~A.}\ \bibnamefont
  {Benalcazar}}, \bibinfo {author} {\bibfnamefont {B.~A.}\ \bibnamefont
  {Bernevig}}, \ and\ \bibinfo {author} {\bibfnamefont {T.~L.}\ \bibnamefont
  {Hughes}},\ }\href {\doibase 10.1103/PhysRevB.96.245115} {\bibfield
  {journal} {\bibinfo  {journal} {Phys. Rev. B}\ }\textbf {\bibinfo {volume}
  {96}},\ \bibinfo {pages} {245115} (\bibinfo {year}
  {2017}{\natexlab{b}})}\BibitemShut {NoStop}%
\bibitem [{\citenamefont {Langbehn}\ \emph {et~al.}(2017)\citenamefont
  {Langbehn}, \citenamefont {Peng}, \citenamefont {Trifunovic}, \citenamefont
  {von Oppen},\ and\ \citenamefont {Brouwer}}]{Langbehn2017}%
  \BibitemOpen
  \bibfield  {author} {\bibinfo {author} {\bibfnamefont {J.}~\bibnamefont
  {Langbehn}}, \bibinfo {author} {\bibfnamefont {Y.}~\bibnamefont {Peng}},
  \bibinfo {author} {\bibfnamefont {L.}~\bibnamefont {Trifunovic}}, \bibinfo
  {author} {\bibfnamefont {F.}~\bibnamefont {von Oppen}}, \ and\ \bibinfo
  {author} {\bibfnamefont {P.~W.}\ \bibnamefont {Brouwer}},\ }\href {\doibase
  10.1103/PhysRevLett.119.246401} {\bibfield  {journal} {\bibinfo  {journal}
  {Phys. Rev. Lett.}\ }\textbf {\bibinfo {volume} {119}},\ \bibinfo {pages}
  {246401} (\bibinfo {year} {2017})}\BibitemShut {NoStop}%
\bibitem [{\citenamefont {Ezawa}(2018)}]{Ezawa2018}%
  \BibitemOpen
  \bibfield  {author} {\bibinfo {author} {\bibfnamefont {M.}~\bibnamefont
  {Ezawa}},\ }\href {\doibase 10.1103/PhysRevLett.120.026801} {\bibfield
  {journal} {\bibinfo  {journal} {Phys. Rev. Lett.}\ }\textbf {\bibinfo
  {volume} {120}},\ \bibinfo {pages} {026801} (\bibinfo {year}
  {2018})}\BibitemShut {NoStop}%
\bibitem [{\citenamefont {Schindler}\ \emph
  {et~al.}(2018{\natexlab{a}})\citenamefont {Schindler}, \citenamefont {Cook},
  \citenamefont {Vergniory}, \citenamefont {Wang}, \citenamefont {Parkin},
  \citenamefont {Bernevig},\ and\ \citenamefont {Neupert}}]{Schindler2018}%
  \BibitemOpen
  \bibfield  {author} {\bibinfo {author} {\bibfnamefont {F.}~\bibnamefont
  {Schindler}}, \bibinfo {author} {\bibfnamefont {A.~M.}\ \bibnamefont {Cook}},
  \bibinfo {author} {\bibfnamefont {M.~G.}\ \bibnamefont {Vergniory}}, \bibinfo
  {author} {\bibfnamefont {Z.}~\bibnamefont {Wang}}, \bibinfo {author}
  {\bibfnamefont {S.~S.~P.}\ \bibnamefont {Parkin}}, \bibinfo {author}
  {\bibfnamefont {B.~A.}\ \bibnamefont {Bernevig}}, \ and\ \bibinfo {author}
  {\bibfnamefont {T.}~\bibnamefont {Neupert}},\ }\href {\doibase
  10.1126/sciadv.aat0346} {\bibfield  {journal} {\bibinfo  {journal} {Science
  Advances}\ }\textbf {\bibinfo {volume} {4}} (\bibinfo {year}
  {2018}{\natexlab{a}}),\ 10.1126/sciadv.aat0346}\BibitemShut {NoStop}%
\bibitem [{\citenamefont {Song}\ \emph {et~al.}(2017)\citenamefont {Song},
  \citenamefont {Fang},\ and\ \citenamefont {Fang}}]{Song2017}%
  \BibitemOpen
  \bibfield  {author} {\bibinfo {author} {\bibfnamefont {Z.}~\bibnamefont
  {Song}}, \bibinfo {author} {\bibfnamefont {Z.}~\bibnamefont {Fang}}, \ and\
  \bibinfo {author} {\bibfnamefont {C.}~\bibnamefont {Fang}},\ }\href {\doibase
  10.1103/PhysRevLett.119.246402} {\bibfield  {journal} {\bibinfo  {journal}
  {Phys. Rev. Lett.}\ }\textbf {\bibinfo {volume} {119}},\ \bibinfo {pages}
  {246402} (\bibinfo {year} {2017})}\BibitemShut {NoStop}%
\bibitem [{\citenamefont {Khalaf}\ \emph {et~al.}(2017)\citenamefont {Khalaf},
  \citenamefont {Po}, \citenamefont {Vishwanath},\ and\ \citenamefont
  {Watanabe}}]{Khalaf2017}%
  \BibitemOpen
  \bibfield  {author} {\bibinfo {author} {\bibfnamefont {E.}~\bibnamefont
  {Khalaf}}, \bibinfo {author} {\bibfnamefont {H.~C.}\ \bibnamefont {Po}},
  \bibinfo {author} {\bibfnamefont {A.}~\bibnamefont {Vishwanath}}, \ and\
  \bibinfo {author} {\bibfnamefont {H.}~\bibnamefont {Watanabe}},\ }\href@noop
  {} {\bibfield  {journal} {\bibinfo  {journal} {arXiv:1711.11589}\ } (\bibinfo
  {year} {2017})}\BibitemShut {NoStop}%
\bibitem [{\citenamefont {Khalaf}(2018)}]{Khalaf2018}%
  \BibitemOpen
  \bibfield  {author} {\bibinfo {author} {\bibfnamefont {E.}~\bibnamefont
  {Khalaf}},\ }\href {\doibase 10.1103/PhysRevB.97.205136} {\bibfield
  {journal} {\bibinfo  {journal} {Phys. Rev. B}\ }\textbf {\bibinfo {volume}
  {97}},\ \bibinfo {pages} {205136} (\bibinfo {year} {2018})}\BibitemShut
  {NoStop}%
\bibitem [{\citenamefont {Fang}\ and\ \citenamefont {Fu}(2017)}]{Fang2017}%
  \BibitemOpen
  \bibfield  {author} {\bibinfo {author} {\bibfnamefont {C.}~\bibnamefont
  {Fang}}\ and\ \bibinfo {author} {\bibfnamefont {L.}~\bibnamefont {Fu}},\
  }\href@noop {} {\bibfield  {journal} {\bibinfo  {journal} {arXiv:1709.01929}\
  } (\bibinfo {year} {2017})}\BibitemShut {NoStop}%
\bibitem [{\citenamefont {Zhu}(2018)}]{Zhu2018}%
  \BibitemOpen
  \bibfield  {author} {\bibinfo {author} {\bibfnamefont {X.}~\bibnamefont
  {Zhu}},\ }\href {\doibase 10.1103/PhysRevB.97.205134} {\bibfield  {journal}
  {\bibinfo  {journal} {Phys. Rev. B}\ }\textbf {\bibinfo {volume} {97}},\
  \bibinfo {pages} {205134} (\bibinfo {year} {2018})}\BibitemShut {NoStop}%
\bibitem [{\citenamefont {Yan}\ \emph {et~al.}(2018)\citenamefont {Yan},
  \citenamefont {Song},\ and\ \citenamefont {Wang}}]{Yan2018}%
  \BibitemOpen
  \bibfield  {author} {\bibinfo {author} {\bibfnamefont {Z.}~\bibnamefont
  {Yan}}, \bibinfo {author} {\bibfnamefont {F.}~\bibnamefont {Song}}, \ and\
  \bibinfo {author} {\bibfnamefont {Z.}~\bibnamefont {Wang}},\ }\href@noop {}
  {\bibfield  {journal} {\bibinfo  {journal} {arXiv:1803.08545}\ } (\bibinfo
  {year} {2018})}\BibitemShut {NoStop}%
\bibitem [{\citenamefont {R\"uegg}\ \emph {et~al.}(2013)\citenamefont
  {R\"uegg}, \citenamefont {Coh},\ and\ \citenamefont {Moore}}]{Ruegg2013pra}%
  \BibitemOpen
  \bibfield  {author} {\bibinfo {author} {\bibfnamefont {A.}~\bibnamefont
  {R\"uegg}}, \bibinfo {author} {\bibfnamefont {S.}~\bibnamefont {Coh}}, \ and\
  \bibinfo {author} {\bibfnamefont {J.~E.}\ \bibnamefont {Moore}},\ }\href
  {\doibase 10.1103/PhysRevB.88.155127} {\bibfield  {journal} {\bibinfo
  {journal} {Phys. Rev. B}\ }\textbf {\bibinfo {volume} {88}},\ \bibinfo
  {pages} {155127} (\bibinfo {year} {2013})}\BibitemShut {NoStop}%
\bibitem [{\citenamefont {Haldane}(1988)}]{Haldane1988}%
  \BibitemOpen
  \bibfield  {author} {\bibinfo {author} {\bibfnamefont {F.~D.~M.}\
  \bibnamefont {Haldane}},\ }\href {\doibase 10.1103/PhysRevLett.61.2015}
  {\bibfield  {journal} {\bibinfo  {journal} {Phys. Rev. Lett.}\ }\textbf
  {\bibinfo {volume} {61}},\ \bibinfo {pages} {2015} (\bibinfo {year}
  {1988})}\BibitemShut {NoStop}%
\bibitem [{\citenamefont {R\"uegg}\ and\ \citenamefont
  {Lin}(2013)}]{Ruegg2013prl}%
  \BibitemOpen
  \bibfield  {author} {\bibinfo {author} {\bibfnamefont {A.}~\bibnamefont
  {R\"uegg}}\ and\ \bibinfo {author} {\bibfnamefont {C.}~\bibnamefont {Lin}},\
  }\href {\doibase 10.1103/PhysRevLett.110.046401} {\bibfield  {journal}
  {\bibinfo  {journal} {Phys. Rev. Lett.}\ }\textbf {\bibinfo {volume} {110}},\
  \bibinfo {pages} {046401} (\bibinfo {year} {2013})}\BibitemShut {NoStop}%
\bibitem [{\citenamefont {Huang}\ \emph {et~al.}(2018)\citenamefont {Huang},
  \citenamefont {Kim}, \citenamefont {Efimkin}, \citenamefont {Lovorn},
  \citenamefont {Taniguchi}, \citenamefont {Watanabe}, \citenamefont
  {MacDonald}, \citenamefont {Tutuc},\ and\ \citenamefont {LeRoy}}]{Huang2018}%
  \BibitemOpen
  \bibfield  {author} {\bibinfo {author} {\bibfnamefont {S.}~\bibnamefont
  {Huang}}, \bibinfo {author} {\bibfnamefont {K.}~\bibnamefont {Kim}}, \bibinfo
  {author} {\bibfnamefont {D.~K.}\ \bibnamefont {Efimkin}}, \bibinfo {author}
  {\bibfnamefont {T.}~\bibnamefont {Lovorn}}, \bibinfo {author} {\bibfnamefont
  {T.}~\bibnamefont {Taniguchi}}, \bibinfo {author} {\bibfnamefont
  {K.}~\bibnamefont {Watanabe}}, \bibinfo {author} {\bibfnamefont {A.~H.}\
  \bibnamefont {MacDonald}}, \bibinfo {author} {\bibfnamefont {E.}~\bibnamefont
  {Tutuc}}, \ and\ \bibinfo {author} {\bibfnamefont {B.~J.}\ \bibnamefont
  {LeRoy}},\ }\href {\doibase 10.1103/PhysRevLett.121.037702} {\bibfield
  {journal} {\bibinfo  {journal} {Phys. Rev. Lett.}\ }\textbf {\bibinfo
  {volume} {121}},\ \bibinfo {pages} {037702} (\bibinfo {year}
  {2018})}\BibitemShut {NoStop}%
\bibitem [{\citenamefont {Wu}\ \emph {et~al.}(2018)\citenamefont {Wu},
  \citenamefont {Jian},\ and\ \citenamefont {Xu}}]{Wu2018}%
  \BibitemOpen
  \bibfield  {author} {\bibinfo {author} {\bibfnamefont {X.-C.}\ \bibnamefont
  {Wu}}, \bibinfo {author} {\bibfnamefont {C.-M.}\ \bibnamefont {Jian}}, \ and\
  \bibinfo {author} {\bibfnamefont {C.}~\bibnamefont {Xu}},\ }\href@noop {}
  {\bibfield  {journal} {\bibinfo  {journal} {arXiv:1811.08442}\ } (\bibinfo
  {year} {2018})}\BibitemShut {NoStop}%
\bibitem [{\citenamefont {Ju}\ \emph {et~al.}(2015)\citenamefont {Ju},
  \citenamefont {Shi}, \citenamefont {Nair}, \citenamefont {Lv}, \citenamefont
  {Jin}, \citenamefont {Velasco~Jr}, \citenamefont {Ojeda-Aristizabal},
  \citenamefont {Bechtel}, \citenamefont {Martin}, \citenamefont {Zettl},
  \citenamefont {Analytis},\ and\ \citenamefont {Wang}}]{fengwang}%
  \BibitemOpen
  \bibfield  {author} {\bibinfo {author} {\bibfnamefont {L.}~\bibnamefont
  {Ju}}, \bibinfo {author} {\bibfnamefont {Z.}~\bibnamefont {Shi}}, \bibinfo
  {author} {\bibfnamefont {N.}~\bibnamefont {Nair}}, \bibinfo {author}
  {\bibfnamefont {Y.}~\bibnamefont {Lv}}, \bibinfo {author} {\bibfnamefont
  {C.}~\bibnamefont {Jin}}, \bibinfo {author} {\bibfnamefont {J.}~\bibnamefont
  {Velasco~Jr}}, \bibinfo {author} {\bibfnamefont {C.}~\bibnamefont
  {Ojeda-Aristizabal}}, \bibinfo {author} {\bibfnamefont {H.~A.}\ \bibnamefont
  {Bechtel}}, \bibinfo {author} {\bibfnamefont {M.~C.}\ \bibnamefont {Martin}},
  \bibinfo {author} {\bibfnamefont {A.}~\bibnamefont {Zettl}}, \bibinfo
  {author} {\bibfnamefont {J.}~\bibnamefont {Analytis}}, \ and\ \bibinfo
  {author} {\bibfnamefont {F.}~\bibnamefont {Wang}},\ }\href@noop {} {\bibfield
   {journal} {\bibinfo  {journal} {Nature}\ }\textbf {\bibinfo {volume}
  {520}},\ \bibinfo {pages} {650 EP } (\bibinfo {year} {2015})}\BibitemShut
  {NoStop}%
\bibitem [{\citenamefont {Teo}\ and\ \citenamefont {Hughes}(2013)}]{Teo2013}%
  \BibitemOpen
  \bibfield  {author} {\bibinfo {author} {\bibfnamefont {J.~C.~Y.}\
  \bibnamefont {Teo}}\ and\ \bibinfo {author} {\bibfnamefont {T.~L.}\
  \bibnamefont {Hughes}},\ }\href {\doibase 10.1103/PhysRevLett.111.047006}
  {\bibfield  {journal} {\bibinfo  {journal} {Phys. Rev. Lett.}\ }\textbf
  {\bibinfo {volume} {111}},\ \bibinfo {pages} {047006} (\bibinfo {year}
  {2013})}\BibitemShut {NoStop}%
\bibitem [{\citenamefont {Gopalakrishnan}\ \emph {et~al.}(2013)\citenamefont
  {Gopalakrishnan}, \citenamefont {Teo},\ and\ \citenamefont
  {Hughes}}]{Gopalakrishnan2013}%
  \BibitemOpen
  \bibfield  {author} {\bibinfo {author} {\bibfnamefont {S.}~\bibnamefont
  {Gopalakrishnan}}, \bibinfo {author} {\bibfnamefont {J.~C.~Y.}\ \bibnamefont
  {Teo}}, \ and\ \bibinfo {author} {\bibfnamefont {T.~L.}\ \bibnamefont
  {Hughes}},\ }\href {\doibase 10.1103/PhysRevLett.111.025304} {\bibfield
  {journal} {\bibinfo  {journal} {Phys. Rev. Lett.}\ }\textbf {\bibinfo
  {volume} {111}},\ \bibinfo {pages} {025304} (\bibinfo {year}
  {2013})}\BibitemShut {NoStop}%
\bibitem [{\citenamefont {Benalcazar}\ \emph {et~al.}(2014)\citenamefont
  {Benalcazar}, \citenamefont {Teo},\ and\ \citenamefont
  {Hughes}}]{Benalcazar2014}%
  \BibitemOpen
  \bibfield  {author} {\bibinfo {author} {\bibfnamefont {W.~A.}\ \bibnamefont
  {Benalcazar}}, \bibinfo {author} {\bibfnamefont {J.~C.~Y.}\ \bibnamefont
  {Teo}}, \ and\ \bibinfo {author} {\bibfnamefont {T.~L.}\ \bibnamefont
  {Hughes}},\ }\href {\doibase 10.1103/PhysRevB.89.224503} {\bibfield
  {journal} {\bibinfo  {journal} {Phys. Rev. B}\ }\textbf {\bibinfo {volume}
  {89}},\ \bibinfo {pages} {224503} (\bibinfo {year} {2014})}\BibitemShut
  {NoStop}%
\bibitem [{\citenamefont {Teo}\ and\ \citenamefont {Kane}(2010)}]{Teo2010}%
  \BibitemOpen
  \bibfield  {author} {\bibinfo {author} {\bibfnamefont {J.~C.~Y.}\
  \bibnamefont {Teo}}\ and\ \bibinfo {author} {\bibfnamefont {C.~L.}\
  \bibnamefont {Kane}},\ }\href {\doibase 10.1103/PhysRevB.82.115120}
  {\bibfield  {journal} {\bibinfo  {journal} {Phys. Rev. B}\ }\textbf {\bibinfo
  {volume} {82}},\ \bibinfo {pages} {115120} (\bibinfo {year}
  {2010})}\BibitemShut {NoStop}%
\bibitem [{\citenamefont {Chiu}\ \emph {et~al.}(2016)\citenamefont {Chiu},
  \citenamefont {Teo}, \citenamefont {Schnyder},\ and\ \citenamefont
  {Ryu}}]{Chiu2016}%
  \BibitemOpen
  \bibfield  {author} {\bibinfo {author} {\bibfnamefont {C.-K.}\ \bibnamefont
  {Chiu}}, \bibinfo {author} {\bibfnamefont {J.~C.~Y.}\ \bibnamefont {Teo}},
  \bibinfo {author} {\bibfnamefont {A.~P.}\ \bibnamefont {Schnyder}}, \ and\
  \bibinfo {author} {\bibfnamefont {S.}~\bibnamefont {Ryu}},\ }\href {\doibase
  10.1103/RevModPhys.88.035005} {\bibfield  {journal} {\bibinfo  {journal}
  {Rev. Mod. Phys.}\ }\textbf {\bibinfo {volume} {88}},\ \bibinfo {pages}
  {035005} (\bibinfo {year} {2016})}\BibitemShut {NoStop}%
\bibitem [{\citenamefont {Lee}\ \emph {et~al.}(2007)\citenamefont {Lee},
  \citenamefont {Zhang},\ and\ \citenamefont {Xiang}}]{Lee2007}%
  \BibitemOpen
  \bibfield  {author} {\bibinfo {author} {\bibfnamefont {D.-H.}\ \bibnamefont
  {Lee}}, \bibinfo {author} {\bibfnamefont {G.-M.}\ \bibnamefont {Zhang}}, \
  and\ \bibinfo {author} {\bibfnamefont {T.}~\bibnamefont {Xiang}},\ }\href
  {\doibase 10.1103/PhysRevLett.99.196805} {\bibfield  {journal} {\bibinfo
  {journal} {Phys. Rev. Lett.}\ }\textbf {\bibinfo {volume} {99}},\ \bibinfo
  {pages} {196805} (\bibinfo {year} {2007})}\BibitemShut {NoStop}%
\bibitem [{\citenamefont {Qi}\ \emph {et~al.}(2008)\citenamefont {Qi},
  \citenamefont {Hughes},\ and\ \citenamefont {Zhang}}]{Qi2008}%
  \BibitemOpen
  \bibfield  {author} {\bibinfo {author} {\bibfnamefont {X.-L.}\ \bibnamefont
  {Qi}}, \bibinfo {author} {\bibfnamefont {T.~L.}\ \bibnamefont {Hughes}}, \
  and\ \bibinfo {author} {\bibfnamefont {S.-C.}\ \bibnamefont {Zhang}},\ }\href
  {\doibase 10.1038/nphys913} {\bibfield  {journal} {\bibinfo  {journal}
  {Nature Physics}\ }\textbf {\bibinfo {volume} {4}},\ \bibinfo {pages} {273}
  (\bibinfo {year} {2008})}\BibitemShut {NoStop}%
\bibitem [{\citenamefont {Ran}\ \emph {et~al.}(2009)\citenamefont {Ran},
  \citenamefont {Zhang},\ and\ \citenamefont {Vishwanath}}]{Ran2009}%
  \BibitemOpen
  \bibfield  {author} {\bibinfo {author} {\bibfnamefont {Y.}~\bibnamefont
  {Ran}}, \bibinfo {author} {\bibfnamefont {Y.}~\bibnamefont {Zhang}}, \ and\
  \bibinfo {author} {\bibfnamefont {A.}~\bibnamefont {Vishwanath}},\ }\href
  {\doibase 10.1038/nphys1220} {\bibfield  {journal} {\bibinfo  {journal}
  {Nature Physics}\ }\textbf {\bibinfo {volume} {5}},\ \bibinfo {pages} {298}
  (\bibinfo {year} {2009})}\BibitemShut {NoStop}%
\bibitem [{\citenamefont {Klinovaja}\ and\ \citenamefont
  {Loss}(2015)}]{Klinovaja2015}%
  \BibitemOpen
  \bibfield  {author} {\bibinfo {author} {\bibfnamefont {J.}~\bibnamefont
  {Klinovaja}}\ and\ \bibinfo {author} {\bibfnamefont {D.}~\bibnamefont
  {Loss}},\ }\href {\doibase 10.1103/PhysRevB.92.121410} {\bibfield  {journal}
  {\bibinfo  {journal} {Phys. Rev. B}\ }\textbf {\bibinfo {volume} {92}},\
  \bibinfo {pages} {121410} (\bibinfo {year} {2015})}\BibitemShut {NoStop}%
\bibitem [{\citenamefont {Su}\ \emph {et~al.}(1980)\citenamefont {Su},
  \citenamefont {Schrieffer},\ and\ \citenamefont {Heeger}}]{Su1980}%
  \BibitemOpen
  \bibfield  {author} {\bibinfo {author} {\bibfnamefont {W.~P.}\ \bibnamefont
  {Su}}, \bibinfo {author} {\bibfnamefont {J.~R.}\ \bibnamefont {Schrieffer}},
  \ and\ \bibinfo {author} {\bibfnamefont {A.~J.}\ \bibnamefont {Heeger}},\
  }\href {\doibase 10.1103/PhysRevB.22.2099} {\bibfield  {journal} {\bibinfo
  {journal} {Phys. Rev. B}\ }\textbf {\bibinfo {volume} {22}},\ \bibinfo
  {pages} {2099} (\bibinfo {year} {1980})}\BibitemShut {NoStop}%
\bibitem [{\citenamefont {Goldstone}\ and\ \citenamefont
  {Wilczek}(1981)}]{Goldstone1981}%
  \BibitemOpen
  \bibfield  {author} {\bibinfo {author} {\bibfnamefont {J.}~\bibnamefont
  {Goldstone}}\ and\ \bibinfo {author} {\bibfnamefont {F.}~\bibnamefont
  {Wilczek}},\ }\href {\doibase 10.1103/PhysRevLett.47.986} {\bibfield
  {journal} {\bibinfo  {journal} {Phys. Rev. Lett.}\ }\textbf {\bibinfo
  {volume} {47}},\ \bibinfo {pages} {986} (\bibinfo {year} {1981})}\BibitemShut
  {NoStop}%
\bibitem [{\citenamefont {Jackiw}\ and\ \citenamefont
  {Semenoff}(1983)}]{Jackiw1983}%
  \BibitemOpen
  \bibfield  {author} {\bibinfo {author} {\bibfnamefont {R.}~\bibnamefont
  {Jackiw}}\ and\ \bibinfo {author} {\bibfnamefont {G.}~\bibnamefont
  {Semenoff}},\ }\href {\doibase 10.1103/PhysRevLett.50.439} {\bibfield
  {journal} {\bibinfo  {journal} {Phys. Rev. Lett.}\ }\textbf {\bibinfo
  {volume} {50}},\ \bibinfo {pages} {439} (\bibinfo {year} {1983})}\BibitemShut
  {NoStop}%
\bibitem [{\citenamefont {Wu}\ \emph {et~al.}(2006)\citenamefont {Wu},
  \citenamefont {Bernevig},\ and\ \citenamefont {Zhang}}]{Wu2006}%
  \BibitemOpen
  \bibfield  {author} {\bibinfo {author} {\bibfnamefont {C.}~\bibnamefont
  {Wu}}, \bibinfo {author} {\bibfnamefont {B.~A.}\ \bibnamefont {Bernevig}}, \
  and\ \bibinfo {author} {\bibfnamefont {S.-C.}\ \bibnamefont {Zhang}},\ }\href
  {\doibase 10.1103/PhysRevLett.96.106401} {\bibfield  {journal} {\bibinfo
  {journal} {Phys. Rev. Lett.}\ }\textbf {\bibinfo {volume} {96}},\ \bibinfo
  {pages} {106401} (\bibinfo {year} {2006})}\BibitemShut {NoStop}%
\bibitem [{\citenamefont {Xu}\ and\ \citenamefont {Moore}(2006)}]{Xu2006}%
  \BibitemOpen
  \bibfield  {author} {\bibinfo {author} {\bibfnamefont {C.}~\bibnamefont
  {Xu}}\ and\ \bibinfo {author} {\bibfnamefont {J.~E.}\ \bibnamefont {Moore}},\
  }\href {\doibase 10.1103/PhysRevB.73.045322} {\bibfield  {journal} {\bibinfo
  {journal} {Phys. Rev. B}\ }\textbf {\bibinfo {volume} {73}},\ \bibinfo
  {pages} {045322} (\bibinfo {year} {2006})}\BibitemShut {NoStop}%
\bibitem [{\citenamefont {Hou}\ \emph {et~al.}(2009)\citenamefont {Hou},
  \citenamefont {Kim},\ and\ \citenamefont {Chamon}}]{Hou2009}%
  \BibitemOpen
  \bibfield  {author} {\bibinfo {author} {\bibfnamefont {C.-Y.}\ \bibnamefont
  {Hou}}, \bibinfo {author} {\bibfnamefont {E.-A.}\ \bibnamefont {Kim}}, \ and\
  \bibinfo {author} {\bibfnamefont {C.}~\bibnamefont {Chamon}},\ }\href
  {\doibase 10.1103/PhysRevLett.102.076602} {\bibfield  {journal} {\bibinfo
  {journal} {Phys. Rev. Lett.}\ }\textbf {\bibinfo {volume} {102}},\ \bibinfo
  {pages} {076602} (\bibinfo {year} {2009})}\BibitemShut {NoStop}%
\bibitem [{\citenamefont {Giamarchi}()}]{Giamarchi2003quantum}%
  \BibitemOpen
  \bibfield  {author} {\bibinfo {author} {\bibfnamefont {T.}~\bibnamefont
  {Giamarchi}},\ }\href@noop {} {\emph {\bibinfo {title} {Quantum Physics in
  One Dimension}}},\ International Series of Monographs on Physics\BibitemShut
  {NoStop}%
\bibitem [{\citenamefont {Chamon}\ \emph {et~al.}()\citenamefont {Chamon},
  \citenamefont {Goerbig}, \citenamefont {Moessner},\ and\ \citenamefont
  {Cugliandolo}}]{Chamon2017topological}%
  \BibitemOpen
  \bibfield  {author} {\bibinfo {author} {\bibfnamefont {C.}~\bibnamefont
  {Chamon}}, \bibinfo {author} {\bibfnamefont {M.}~\bibnamefont {Goerbig}},
  \bibinfo {author} {\bibfnamefont {R.}~\bibnamefont {Moessner}}, \ and\
  \bibinfo {author} {\bibfnamefont {L.}~\bibnamefont {Cugliandolo}},\
  }\href@noop {} {\emph {\bibinfo {title} {Topological Aspects of Condensed
  Matter Physics: Lecture Notes of the Les Houches Summer School: Volume 103,
  August 2014}}},\ Lecture Notes of the Les Houches Summer School\BibitemShut
  {NoStop}%
\bibitem [{\citenamefont {Shankar}(2017)}]{Shankar2017}%
  \BibitemOpen
  \bibfield  {author} {\bibinfo {author} {\bibfnamefont {R.}~\bibnamefont
  {Shankar}},\ }\href {\doibase 10.1017/9781139044349.018} {\emph {\bibinfo
  {title} {Quantum Field Theory and Condensed Matter: An Introduction}}}\
  (\bibinfo  {publisher} {Cambridge University Press},\ \bibinfo {year}
  {2017})\ pp.\ \bibinfo {pages} {319--333}\BibitemShut {NoStop}%
\bibitem [{\citenamefont {Fr{\"o}hlich}\ and\ \citenamefont
  {Marchetti}(1988)}]{Frohlich1988}%
  \BibitemOpen
  \bibfield  {author} {\bibinfo {author} {\bibfnamefont {J.}~\bibnamefont
  {Fr{\"o}hlich}}\ and\ \bibinfo {author} {\bibfnamefont {P.}~\bibnamefont
  {Marchetti}},\ }\href {\doibase 10.1007/BF01239028} {\bibfield  {journal}
  {\bibinfo  {journal} {Communications in Mathematical Physics}\ }\textbf
  {\bibinfo {volume} {116}},\ \bibinfo {pages} {127} (\bibinfo {year}
  {1988})}\BibitemShut {NoStop}%
\bibitem [{\citenamefont {Oshikawa}\ \emph {et~al.}(2006)\citenamefont
  {Oshikawa}, \citenamefont {Chamon},\ and\ \citenamefont
  {Affleck}}]{affleckoshikawa}%
  \BibitemOpen
  \bibfield  {author} {\bibinfo {author} {\bibfnamefont {M.}~\bibnamefont
  {Oshikawa}}, \bibinfo {author} {\bibfnamefont {C.}~\bibnamefont {Chamon}}, \
  and\ \bibinfo {author} {\bibfnamefont {I.}~\bibnamefont {Affleck}},\ }\href
  {http://stacks.iop.org/1742-5468/2006/i=02/a=P02008} {\bibfield  {journal}
  {\bibinfo  {journal} {Journal of Statistical Mechanics: Theory and
  Experiment}\ }\textbf {\bibinfo {volume} {2006}},\ \bibinfo {pages} {P02008}
  (\bibinfo {year} {2006})}\BibitemShut {NoStop}%
\bibitem [{\citenamefont {Hou}\ \emph {et~al.}(2012)\citenamefont {Hou},
  \citenamefont {Rahmani}, \citenamefont {Feiguin},\ and\ \citenamefont
  {Chamon}}]{Hou2012}%
  \BibitemOpen
  \bibfield  {author} {\bibinfo {author} {\bibfnamefont {C.-Y.}\ \bibnamefont
  {Hou}}, \bibinfo {author} {\bibfnamefont {A.}~\bibnamefont {Rahmani}},
  \bibinfo {author} {\bibfnamefont {A.~E.}\ \bibnamefont {Feiguin}}, \ and\
  \bibinfo {author} {\bibfnamefont {C.}~\bibnamefont {Chamon}},\ }\href
  {\doibase 10.1103/PhysRevB.86.075451} {\bibfield  {journal} {\bibinfo
  {journal} {Phys. Rev. B}\ }\textbf {\bibinfo {volume} {86}},\ \bibinfo
  {pages} {075451} (\bibinfo {year} {2012})}\BibitemShut {NoStop}%
\bibitem [{\citenamefont {Shiozaki}\ and\ \citenamefont
  {Sato}(2014)}]{Shiozaki2014}%
  \BibitemOpen
  \bibfield  {author} {\bibinfo {author} {\bibfnamefont {K.}~\bibnamefont
  {Shiozaki}}\ and\ \bibinfo {author} {\bibfnamefont {M.}~\bibnamefont
  {Sato}},\ }\href {\doibase 10.1103/PhysRevB.90.165114} {\bibfield  {journal}
  {\bibinfo  {journal} {Phys. Rev. B}\ }\textbf {\bibinfo {volume} {90}},\
  \bibinfo {pages} {165114} (\bibinfo {year} {2014})}\BibitemShut {NoStop}%
\bibitem [{\citenamefont {Lau}\ \emph {et~al.}(2016)\citenamefont {Lau},
  \citenamefont {van~den Brink},\ and\ \citenamefont {Ortix}}]{Lau2016}%
  \BibitemOpen
  \bibfield  {author} {\bibinfo {author} {\bibfnamefont {A.}~\bibnamefont
  {Lau}}, \bibinfo {author} {\bibfnamefont {J.}~\bibnamefont {van~den Brink}},
  \ and\ \bibinfo {author} {\bibfnamefont {C.}~\bibnamefont {Ortix}},\ }\href
  {\doibase 10.1103/PhysRevB.94.165164} {\bibfield  {journal} {\bibinfo
  {journal} {Phys. Rev. B}\ }\textbf {\bibinfo {volume} {94}},\ \bibinfo
  {pages} {165164} (\bibinfo {year} {2016})}\BibitemShut {NoStop}%
\bibitem [{\citenamefont {Trifunovic}\ and\ \citenamefont
  {Brouwer}(2017)}]{Trifunovic2017}%
  \BibitemOpen
  \bibfield  {author} {\bibinfo {author} {\bibfnamefont {L.}~\bibnamefont
  {Trifunovic}}\ and\ \bibinfo {author} {\bibfnamefont {P.}~\bibnamefont
  {Brouwer}},\ }\href {\doibase 10.1103/PhysRevB.96.195109} {\bibfield
  {journal} {\bibinfo  {journal} {Phys. Rev. B}\ }\textbf {\bibinfo {volume}
  {96}},\ \bibinfo {pages} {195109} (\bibinfo {year} {2017})}\BibitemShut
  {NoStop}%
\bibitem [{\citenamefont {Serra-Garcia}\ \emph {et~al.}(2018)\citenamefont
  {Serra-Garcia}, \citenamefont {Peri}, \citenamefont {S{\"u}sstrunk},
  \citenamefont {Bilal}, \citenamefont {Larsen}, \citenamefont {Villanueva},\
  and\ \citenamefont {Huber}}]{Serra-Garcia2018}%
  \BibitemOpen
  \bibfield  {author} {\bibinfo {author} {\bibfnamefont {M.}~\bibnamefont
  {Serra-Garcia}}, \bibinfo {author} {\bibfnamefont {V.}~\bibnamefont {Peri}},
  \bibinfo {author} {\bibfnamefont {R.}~\bibnamefont {S{\"u}sstrunk}}, \bibinfo
  {author} {\bibfnamefont {O.~R.}\ \bibnamefont {Bilal}}, \bibinfo {author}
  {\bibfnamefont {T.}~\bibnamefont {Larsen}}, \bibinfo {author} {\bibfnamefont
  {L.~G.}\ \bibnamefont {Villanueva}}, \ and\ \bibinfo {author} {\bibfnamefont
  {S.~D.}\ \bibnamefont {Huber}},\ }\href {\doibase 10.1038/nature25156}
  {\bibfield  {journal} {\bibinfo  {journal} {Nature}\ }\textbf {\bibinfo
  {volume} {555}},\ \bibinfo {pages} {342 } (\bibinfo {year}
  {2018})}\BibitemShut {NoStop}%
\bibitem [{\citenamefont {Peterson}\ \emph {et~al.}(2018)\citenamefont
  {Peterson}, \citenamefont {Benalcazar}, \citenamefont {Hughes},\ and\
  \citenamefont {Bahl}}]{Peterson2018}%
  \BibitemOpen
  \bibfield  {author} {\bibinfo {author} {\bibfnamefont {C.~W.}\ \bibnamefont
  {Peterson}}, \bibinfo {author} {\bibfnamefont {W.~A.}\ \bibnamefont
  {Benalcazar}}, \bibinfo {author} {\bibfnamefont {T.~L.}\ \bibnamefont
  {Hughes}}, \ and\ \bibinfo {author} {\bibfnamefont {G.}~\bibnamefont
  {Bahl}},\ }\href {\doibase 10.1038/nature25777} {\bibfield  {journal}
  {\bibinfo  {journal} {Nature}\ }\textbf {\bibinfo {volume} {555}},\ \bibinfo
  {pages} {346 } (\bibinfo {year} {2018})}\BibitemShut {NoStop}%
\bibitem [{\citenamefont {Imhof}\ \emph {et~al.}(2017)\citenamefont {Imhof},
  \citenamefont {Berger}, \citenamefont {Bayer}, \citenamefont {Brehm},
  \citenamefont {Molenkamp}, \citenamefont {Kiessling}, \citenamefont
  {Schindler}, \citenamefont {Lee}, \citenamefont {Greiter}, \citenamefont
  {Neupert},\ and\ \citenamefont {Thomale}}]{Imhof2017}%
  \BibitemOpen
  \bibfield  {author} {\bibinfo {author} {\bibfnamefont {S.}~\bibnamefont
  {Imhof}}, \bibinfo {author} {\bibfnamefont {C.}~\bibnamefont {Berger}},
  \bibinfo {author} {\bibfnamefont {F.}~\bibnamefont {Bayer}}, \bibinfo
  {author} {\bibfnamefont {J.}~\bibnamefont {Brehm}}, \bibinfo {author}
  {\bibfnamefont {L.}~\bibnamefont {Molenkamp}}, \bibinfo {author}
  {\bibfnamefont {T.}~\bibnamefont {Kiessling}}, \bibinfo {author}
  {\bibfnamefont {F.}~\bibnamefont {Schindler}}, \bibinfo {author}
  {\bibfnamefont {C.~H.}\ \bibnamefont {Lee}}, \bibinfo {author} {\bibfnamefont
  {M.}~\bibnamefont {Greiter}}, \bibinfo {author} {\bibfnamefont
  {T.}~\bibnamefont {Neupert}}, \ and\ \bibinfo {author} {\bibfnamefont
  {R.}~\bibnamefont {Thomale}},\ }\href@noop {} {\bibfield  {journal} {\bibinfo
   {journal} {arXiv:1708.03647}\ } (\bibinfo {year} {2017})}\BibitemShut
  {NoStop}%
\bibitem [{\citenamefont {Schindler}\ \emph
  {et~al.}(2018{\natexlab{b}})\citenamefont {Schindler}, \citenamefont {Wang},
  \citenamefont {Vergniory}, \citenamefont {Cook}, \citenamefont {Murani},
  \citenamefont {Sengupta}, \citenamefont {Kasumov}, \citenamefont {Deblock},
  \citenamefont {Jeon}, \citenamefont {Drozdov}, \citenamefont {Bouchiat},
  \citenamefont {Gu{\'e}ron}, \citenamefont {Yazdani}, \citenamefont
  {Bernevig},\ and\ \citenamefont {Neupert}}]{Schindler2018Nature}%
  \BibitemOpen
  \bibfield  {author} {\bibinfo {author} {\bibfnamefont {F.}~\bibnamefont
  {Schindler}}, \bibinfo {author} {\bibfnamefont {Z.}~\bibnamefont {Wang}},
  \bibinfo {author} {\bibfnamefont {M.~G.}\ \bibnamefont {Vergniory}}, \bibinfo
  {author} {\bibfnamefont {A.~M.}\ \bibnamefont {Cook}}, \bibinfo {author}
  {\bibfnamefont {A.}~\bibnamefont {Murani}}, \bibinfo {author} {\bibfnamefont
  {S.}~\bibnamefont {Sengupta}}, \bibinfo {author} {\bibfnamefont {A.~Y.}\
  \bibnamefont {Kasumov}}, \bibinfo {author} {\bibfnamefont {R.}~\bibnamefont
  {Deblock}}, \bibinfo {author} {\bibfnamefont {S.}~\bibnamefont {Jeon}},
  \bibinfo {author} {\bibfnamefont {I.}~\bibnamefont {Drozdov}}, \bibinfo
  {author} {\bibfnamefont {H.}~\bibnamefont {Bouchiat}}, \bibinfo {author}
  {\bibfnamefont {S.}~\bibnamefont {Gu{\'e}ron}}, \bibinfo {author}
  {\bibfnamefont {A.}~\bibnamefont {Yazdani}}, \bibinfo {author} {\bibfnamefont
  {B.~A.}\ \bibnamefont {Bernevig}}, \ and\ \bibinfo {author} {\bibfnamefont
  {T.}~\bibnamefont {Neupert}},\ }\href {\doibase 10.1038/s41567-018-0224-7}
  {\bibfield  {journal} {\bibinfo  {journal} {Nature Physics}\ }\textbf
  {\bibinfo {volume} {14}},\ \bibinfo {pages} {918} (\bibinfo {year}
  {2018}{\natexlab{b}})}\BibitemShut {NoStop}%
\bibitem [{\citenamefont {Liu}\ \emph {et~al.}(2014)\citenamefont {Liu},
  \citenamefont {Law},\ and\ \citenamefont {Ng}}]{Liu2014}%
  \BibitemOpen
  \bibfield  {author} {\bibinfo {author} {\bibfnamefont {X.-J.}\ \bibnamefont
  {Liu}}, \bibinfo {author} {\bibfnamefont {K.~T.}\ \bibnamefont {Law}}, \ and\
  \bibinfo {author} {\bibfnamefont {T.~K.}\ \bibnamefont {Ng}},\ }\href
  {\doibase 10.1103/PhysRevLett.112.086401} {\bibfield  {journal} {\bibinfo
  {journal} {Phys. Rev. Lett.}\ }\textbf {\bibinfo {volume} {112}},\ \bibinfo
  {pages} {086401} (\bibinfo {year} {2014})}\BibitemShut {NoStop}%
\bibitem [{\citenamefont {Wu}\ \emph {et~al.}(2016)\citenamefont {Wu},
  \citenamefont {Zhang}, \citenamefont {Sun}, \citenamefont {Xu}, \citenamefont
  {Wang}, \citenamefont {Ji}, \citenamefont {Deng}, \citenamefont {Chen},
  \citenamefont {Liu},\ and\ \citenamefont {Pan}}]{Wu2016}%
  \BibitemOpen
  \bibfield  {author} {\bibinfo {author} {\bibfnamefont {Z.}~\bibnamefont
  {Wu}}, \bibinfo {author} {\bibfnamefont {L.}~\bibnamefont {Zhang}}, \bibinfo
  {author} {\bibfnamefont {W.}~\bibnamefont {Sun}}, \bibinfo {author}
  {\bibfnamefont {X.-T.}\ \bibnamefont {Xu}}, \bibinfo {author} {\bibfnamefont
  {B.-Z.}\ \bibnamefont {Wang}}, \bibinfo {author} {\bibfnamefont {S.-C.}\
  \bibnamefont {Ji}}, \bibinfo {author} {\bibfnamefont {Y.}~\bibnamefont
  {Deng}}, \bibinfo {author} {\bibfnamefont {S.}~\bibnamefont {Chen}}, \bibinfo
  {author} {\bibfnamefont {X.-J.}\ \bibnamefont {Liu}}, \ and\ \bibinfo
  {author} {\bibfnamefont {J.-W.}\ \bibnamefont {Pan}},\ }\href {\doibase
  10.1126/science.aaf6689} {\bibfield  {journal} {\bibinfo  {journal}
  {Science}\ }\textbf {\bibinfo {volume} {354}},\ \bibinfo {pages} {83}
  (\bibinfo {year} {2016})}\BibitemShut {NoStop}%
\bibitem [{\citenamefont {Zhou}\ \emph {et~al.}(2008)\citenamefont {Zhou},
  \citenamefont {Lu}, \citenamefont {Chu}, \citenamefont {Shen},\ and\
  \citenamefont {Niu}}]{Zhou2008}%
  \BibitemOpen
  \bibfield  {author} {\bibinfo {author} {\bibfnamefont {B.}~\bibnamefont
  {Zhou}}, \bibinfo {author} {\bibfnamefont {H.-Z.}\ \bibnamefont {Lu}},
  \bibinfo {author} {\bibfnamefont {R.-L.}\ \bibnamefont {Chu}}, \bibinfo
  {author} {\bibfnamefont {S.-Q.}\ \bibnamefont {Shen}}, \ and\ \bibinfo
  {author} {\bibfnamefont {Q.}~\bibnamefont {Niu}},\ }\href {\doibase
  10.1103/PhysRevLett.101.246807} {\bibfield  {journal} {\bibinfo  {journal}
  {Phys. Rev. Lett.}\ }\textbf {\bibinfo {volume} {101}},\ \bibinfo {pages}
  {246807} (\bibinfo {year} {2008})}\BibitemShut {NoStop}%
\bibitem [{\citenamefont {Bianco}\ and\ \citenamefont
  {Resta}(2011)}]{Bianco2011}%
  \BibitemOpen
  \bibfield  {author} {\bibinfo {author} {\bibfnamefont {R.}~\bibnamefont
  {Bianco}}\ and\ \bibinfo {author} {\bibfnamefont {R.}~\bibnamefont {Resta}},\
  }\href {\doibase 10.1103/PhysRevB.84.241106} {\bibfield  {journal} {\bibinfo
  {journal} {Phys. Rev. B}\ }\textbf {\bibinfo {volume} {84}},\ \bibinfo
  {pages} {241106} (\bibinfo {year} {2011})}\BibitemShut {NoStop}%
\end{thebibliography}%
\noindent

\end{document}